\documentclass[amsmath,amssymb,floatfix,prb,epsf,showpacs,twocolumn,superscriptaddress]{revtex4-2}
\usepackage{standalone}
\usepackage{graphics}
\usepackage{amsmath}
\usepackage{bigints}
\usepackage{amscd}
\usepackage{makeidx}
\usepackage{multirow}
\usepackage{mathtools}
\usepackage{amsfonts}
\usepackage[dvipsnames]{xcolor}
\usepackage{graphicx}
\usepackage{subfig}
\usepackage{siunitx}
\usepackage{booktabs}
\usepackage{xcolor}
\usepackage{tikz}
\usetikzlibrary{3d} 
\usetikzlibrary {arrows.meta}
\usepackage{soul}
\tikzset
{
   axis/.style={thick,-latex},
   yz/.style={canvas is yz plane at x=0},
   xz/.style={canvas is xz plane at y=0},
   xz rotated/.style={rotate around z=\mytheta,xz},
   xy elevated/.style={canvas is xy plane at z={\myradius*cos(\myphi)},scale={sin(\myphi)}},
   sphere/.style={shading=ball,fill opacity=0.3},
   plane/.style={fill=teal,fill opacity=0.3},
   cone/.style={fill=yellow,fill opacity=0.3},
   vector/.style={very thick,-stealth},
   point/.style={shading=ball,ball color=red}
}
\usetikzlibrary{shapes,calc,arrows,through,intersections}
\usetikzlibrary{arrows.meta,calc,shapes,decorations.pathreplacing,decorations.markings, snakes}
\usepackage{caption}
\usepackage[compat=1.1.0]{tikz-feynman}
\usepackage{feynmp-auto}
\usepackage{epic,eepic,epsfig}
\usepackage{dcolumn}
\usepackage{bm}
\usepackage{color}
\usepackage{bbm}
\usepackage{physics}
\newcommand{\nn}{\nonumber}
\newcommand{\bea}{\begin{eqnarray}}
\newcommand{\ena}{\end{eqnarray}}
\usepackage{pgfplots}
\pgfplotsset{compat=newest}

\usepackage{xcolor}
\colorlet{linkequation}{blue}
\usepackage[colorlinks]{hyperref}
\usepackage{cleveref}
\usepackage{afterpage}

\hypersetup{
    colorlinks=true,
    linkcolor=red,
    filecolor=blue,      
    urlcolor=blue,
    citecolor=green,
    pdftitle={Overleaf Example},
    pdfpagemode=FullScreen,
    }
\makeatother

\begin{document}
\setstcolor{red}
\title{
Chiral vortex-line liquid of three-dimensional interacting Bose systems with moat dispersion}

\author{Bahar Jafari-Zadeh} 
\affiliation{Department of Physics, University of Massachusetts, Amherst, Massachusetts 01003, USA}

\author{Chenan Wei} 
\affiliation{Department of Physics, University of Massachusetts, Amherst, Massachusetts 01003, USA}
\affiliation{A.Alikhanyan National Science Laboratory, Br. Alikhanian 2, Yerevan 0036, Armenia}

\author{Tigran A. Sedrakyan}
\affiliation{Department of Physics, University of Massachusetts, Amherst, Massachusetts 01003, USA}

\begin{abstract}
We formulate and investigate a novel quantum state, the Chiral Vortex-Line Liquid (CVLL), emerging in three-dimensional interacting Bose systems exhibiting moat-band dispersions. Such dispersions feature extensive degeneracy along closed manifolds in momentum space, significantly amplifying quantum fluctuations that suppress conventional Bose-Einstein condensation. By extending the two-dimensional Chern-Simons (CS) flux-attachment transformation to three dimensions through a combination of planar CS phases and Jordan-Wigner fermionization along vortex lines, we construct the CVLL state, characterized by preserved rotational $SO(2)$ symmetry, broken time-reversal symmetry, nontrivial vortex-line excitations, and topological gapless edge surface states.
We construct the associated field theory in a curved spatial geometry and analyze the low-energy effective theory of the CVLL state, demonstrating its topological nature.
Using Monte Carlo simulations, we numerically determine the scaling of the chemical potential of the CVLL ground state as a function of boson density for interacting bosons in a cylindrical moat-band geometry and demonstrate that the CVLL phase energetically outcompetes traditional condensate phases at low densities, highlighting its relevance to experimental platforms including frustrated quantum magnets, ultracold atomic gases, excitonic systems, the physics of rotons in $^4$He, and moat regimes in heavy-ion collisions. 
\end{abstract}


\date{\today}

\maketitle

\section{Introduction}
 In systems with a moat dispersion, the minimum energy does not occur at discrete points but along continuous contours in two-dimensional (2D)~\cite{Berg-2012, ARM-2013, CJS-2011, LS-2016, Spielman21, Balents24, sedrakyan-2012,sedrakyan-2014,sedrakyan-2015-1,sedrakyan-2015-2} or surfaces in 3D~\cite{ST-2004, Galitski-2012, Janos-2013, Pisar-2021,roton1961,roton2018,roton2021,roton2010} reciprocal space. This creates a highly degenerate manifold of low-energy single-particle states. Such dispersions can be engineered using spin-orbit coupling, synthetic gauge fields, or by designing specific lattice geometries and hopping integrals. The moat dispersion is also an inherent property of rotons in $^4$He, a class of frustrated quantum magnets, which emerges in heavy-ion collisions.
The degeneracy of the low-energy manifold in moat bands leads to frustration in bosonic systems. Frustration arises because there are infinitely many ways for the bosons to occupy the degenerate states while minimizing the kinetic energy in the presence of interactions. This contrasts with systems where the energy minima occur at isolated points, leading to a unique momentum (or a finite set of momenta) condensation. With such extensive degeneracy, the role of the interactions is now enhanced. This frustration prevents the system from easily settling into an ordered state and can give rise to novel quantum phases, such as spin liquids or supersolids, where the condensation is absent. 

Interestingly, in systems with moat dispersions, frustration can be relieved by fractionalization. In this context, fractionalization refers to the phenomenon where, due to the long-range entanglement, infinitely many constituent bosonic excitations combine to yield emergent quasiparticles with different statistics, such as fermions or anyons. The fractionalization process can lower system energy by allowing the formation of states that better accommodate frustration. For instance, bosons in a frustrated lattice may fractionalize into fermionic quasiparticles, which can fill the degenerate states more efficiently due to the Pauli exclusion principle. In this work, we will show that in 3D, anyonic vortex-lines — extended objects with non-trivial braiding statistics — can emerge, further enriching the system's behavior~\cite{Frohlich-1989}. The emergence of fermions and anyonic strings from bosonic systems is a profound example of how collective excitations can exhibit statistics that differ from those of their constituent particles. 

In the context of fractionalization of excitations, the Chern-Simons (CS) transformation arises naturally and is achieved by attaching magnetic flux tubes to bosons, altering their statistical behavior~\cite{Polyakov-1988, Dunne-1998, Witten-1988}. The CS gauge field mediates this flux attachment, effectively changing the commutation relations of the particles and leading to new emergent phenomena. In (2+1)-dimensional spacetime, the implication of the CS theory is its ability to describe statistical transmutation and topological order, where the low-energy excitations have fractional statistics intermediate between those of bosons and fermions. The topological order represents a new kind of order in quantum matter that extends beyond the conventional symmetry-breaking framework. It is characterized by ground-state degeneracy dependent on the topology of the underlying manifold and the presence of long-range entanglement.

Extending the concept of topological order and the CS term to four-dimensional spacetime is nontrivial because of differences in topology and the behavior of particle statistics in higher dimensions. In (3+1)-dimensional spacetime, particles can typically avoid each other without any topological obstruction, and the concept of anyonic statistics does not directly generalize. However, topological field theories in (3+1) dimensions can still exhibit rich structures that support extended objects such as strings and membranes. For example, higher-form gauge theories involve fields that couple to these extended objects, and their dynamics can lead to generalized forms of topological order. The analogs of the CS term in higher dimensions involve topological terms such as the theta term in Yang-Mills theories~\cite{Witten-1988, GKKS-2017}, which can have significant physical consequences. Weyl semi-metals and axion insulators~\cite{wang-zhang-2013, WTVS-2011, BB-2011} are examples where higher-dimensional topological effects play a crucial role, with the theta term contributing to their electromagnetic responses. Moreover, the exploration of topological quantum field theories in various dimensions aids in classifying possible topological phases of matter and understanding the constraints imposed by gauge invariance and anomaly cancellation~\cite{wen-2013, anomalyQFT-2022, RyanDom-2018}.

Generalizing concepts like statistical transmutation and topological order to four-dimensional spacetime involves overcoming significant theoretical challenges. The absence of non-trivial braid groups for point particles in higher dimensions necessitates considering extended objects and higher-form symmetries. Ensuring gauge invariance and consistency of the theory requires careful construction of the action and consideration of global anomalies. 
The CS theory cannot be directly defined in (3+1)D due to the antisymmetric nature of the wedge products of the gauge field in the CS action. However, recently, there have been new proposals for extending CS theory to (3+1)D~\cite{Chen-Lam-Ma-2022, iCSM_gapped, Palumbo, Sta-Rud-Han-Wil}.

One of the approaches in the literature is based on considering a semi-classical phase space that supports quasi-strings in the bulk state~\cite{Palumbo}. Such an extended object can be viewed as an excitation in (3+1)D, where the point-like quasi-particles reside at the endpoint of the quasi-strings. This theory is supported by defining a two-form current field ($J_{\mu\nu}$) where a quasiparticle current ($J_{\mu}$) resides at the endpoints of the quasi-strings. This paper implements the main (3+1)D BF theory where the quasi-strings live on worldsheets, and the one-form gauge potential ($A_{\mu}$) is replaced by the Kalb-Ramond gauge potential ($\mathcal{B}_{\mu \nu}$). The second approach is established to define (3+1)D fractonic order by stacking infinite layers of the (2+1)D $U(1)$ gauge field with the Lagrangian including both Maxwell and CS terms named infinite Chern-Simons-Maxwell (iCSM) Theory~[\onlinecite{Chen-Lam-Ma-2022}]. Here, one can define a one-form string-like operator in the $z$-direction as an emergent dimension in the continuum limit. The low-energy excitations discussed in the present work will have a similar string-like nature due to the stacking of infinite layers. Refs.~[\onlinecite{Palumbo, Sta-Rud-Han-Wil}] explore the possibility of the emergent CS interaction in (3+1)D within certain materials, such as chiral superconductors and superconducting Weyl semimetals. The proposed Lagrangian, containing an axion coupled to the electromagnetic field, under some assumptions, suggests that the theory supports independent connected layers of (2+1)D CS.

The lattice CS gauge theory\cite{LCS} in condensed matter physics emerges in the moat band lattices, such as the honeycomb lattice with frustrating nearest and next-nearest neighbor XY interactions \cite{sedrakyan-2014,sedrakyan-2015-1,emergent2022} or the XY model on the Kagome lattice supporting a flat band [\onlinecite{distinct2015, Maiti-2019, Bae-2023}]. Here, the 2D lattice version of the flux attachment procedure (lattice CS transformation) allows for the exact reformulation of the frustrated antiferromagnetism problem to the problem of lattice fermions interacting with the CS gauge field~\cite{sedrakyan-2014, sedrakyan-2015-1, emergent2022, MS19, CS-S17, CS-S18, CS-S22,helical20}. The fact that the fermions occupy a Chern band with a nonzero Chern number (which for the Honeycomb lattice is equivalent to the gauged Haldane-Chern insulator [\onlinecite{MS19}]) translates into the fact that the low-energy effective theory is a chiral spin-liquid (CSL) with lattice CS gauge theory with $K=2$. These findings leverage the lattice version of the CS gauge field to describe the topological properties of the CSL phase in flat-band and moat band lattices, overcoming the long-standing challenge of defining a CS theory on a lattice. This formulation of the lattice CS theory has recently been studied in the mathematical physics context in [\onlinecite{Chen-2021, JS-2023, Xu-Chen-2024}]. Ref.~[\onlinecite{Chen-2021}] considers 2D lattices enriched by an electromagnetic $U(1)$ background. It develops a method to describe these phases using lattice models and continuum path integral formulations, revealing novel connections to (twisted) doubled U(1) CS theory through Deligne–Beilinson cohomology. Ref.~[\onlinecite{JS-2023}] offered an approach to regularizing $U(1)$ CS theory on a Euclidean spacetime lattice using a modified Villain framework, which has the potential of supporting global features of continuum CS theory, such as level quantization, 1-form symmetries, and the 't Hooft anomaly directly on the lattice. Along these lines, Ref.~[\onlinecite{Xu-Chen-2024}] addressed two critical challenges in the former approach: the necessity of including a Maxwell term for maintaining local dynamics and implementing the $U(1)$ gauge field in its Villain representation to preserve topological properties, including ground state degeneracy and the chiral gravitational anomaly.

In~\cite{sedrakyan-2012,sedrakyan-2015-1,sedrakyan-2015-2,sedrakyan-2023,sedrakyan-2014,wei2023chiral,emergent2022}, it was shown that in 2D interacting Bose systems with moat-type dispersion, the moat-induced frustration of the system leads to the formation of the CSL phase with semion excitations. The equation of state, namely the scaling of the per-particle ground state energy of such system as a function of the density of Bosons, was shown to be $\sim n^{2}\log^2(n)$, which at low densities is lower than the energy of any bosonic condensate~\cite{GL-2019, WMZ-2011, WGJZ-2010, Ozawa-Baym}. The effective 1D $(\sim1/\sqrt{\epsilon})$ divergence of the density of states (DOS) near the bottom of the moat band is seen in Rashba spin-orbit-coupled electron systems with 2D and 3D moat dispersions~\cite{Capp-Gir-Mars-2007, Elena-Ara, Maiti-Maslov-2015}.
Another interesting study is on the moat-like (1D-like) behavior of the DOS in a 3D topological magnet~\cite{van-hove-2023}. This study observes the 3D Van Hove singularity, the divergence in DOS at critical points in the band, in a 3D topological magnet.  

The 3D moat dispersion is also studied in the context of quantum criticality in a paired fermion system with unbalanced densities~\cite{Yang-Sach-2006}. It was shown that the one-loop polarization operator at small momenta approaches a universal function independent of the repulsive bare interaction. Rashba spin-orbit coupling also leads to the formation of the moat dispersion for electrons. The low-energy sector of Rashba spin-orbit coupled electron systems with interactions in 2D is studied in~\cite{Berg-2012, Berg-2014}. 

Recently, several experiments have studied the many-body physics emerging from the moat dispersion for bosons.
Ref.~\cite{Floquet-Engineered-Moat} presents the experimental realization of a moat-like energy dispersion using Floquet-driven ultracold bosons in an optical lattice, motivated by the theoretical proposal put forward in~[\onlinecite{sedrakyan-2015-2}]. The authors employ Floquet engineering by periodically modulating the depth of a checkerboard optical lattice, hybridizing the two lowest energy bands to create a moat band. The moat-like structure is confirmed by measuring a zero group velocity at a non-zero quasimomentum and investigating how the modified dispersion affects the condensate's center-of-mass trajectory. This work represents a significant step toward achieving strongly correlated quantum states, such as bosonic CSL, within driven optical lattices.

Ref.~\cite{sedrakyan-2023} presents a significant advance in the experimental realization of moat band physics. The study focuses on shallowly inverted InAs/GaSb quantum wells, where the imbalanced electron and hole densities create a frustrated excitonic system. This frustration results in the formation of an excitonic moat band, with energy minima forming a continuous loop in momentum space. The experiments reveal a time-reversal symmetry breaking and the formation of the bosonic CSL, dubbed the excitonic topological order. The observed state remains stable across a range of density imbalances and persists under strong perpendicular magnetic fields up to $35~T$.

Historically, the effect of the moat-like degeneracy for the order parameter in Landau-Ginzburg theory was discussed in Ref.~\cite{Braz-1975}, where the creation of the order parameter of a Bose condensation on the spherical moat in 3D was investigated, focusing on conditions where the order parameter becomes nonuniform at nonzero momenta (condensation at discrete momenta along the moat). It is predicted that under the assumption of condensation, weak interactions, and sufficiently low energies, a uniform condensate can become thermodynamically unstable, leading to a first-order transition towards non-uniform states, such as periodic or helical structures. The study explores scalar and vector order parameters, showing that these transitions can occur without third-order terms in the Landau-Ginzburg functional. These results apply to systems such as cholesteric liquid crystals, rare-earth metals, and magnets with large-period superstructures. The analogous study in 2D was performed in Refs.~\cite{GL-2019, WMZ-2011, WGJZ-2010}. However, Ref.~\cite{wei2023chiral} showed that in 2D,  Bose condensation on the moat bands is ruled out at low densities due to the specific form of the equation of state, where the chemical potential $\mu$ scales as $n^2\log ^2 (n)$ at low particle densities $n\rightarrow 0$. 
This behavior makes the CSL state energetically more favorable compared to the conventional Bose condensation. Through Monte Carlo simulations, Ref.~\cite{wei2023chiral}, the two of us have determined the density range where this scaling law holds, confirming that Bose condensation is suppressed in favor of the CSL state in 2D. 

In the present work, we demonstrate a generalization of the 2D CSL state to the 3D space, dubbed a {\it chiral vortex-line liquid} (CVLL), and present the action formulation of the corresponding state in (3+1)D spacetime. Analogous to the CSL state in 2D, which is based on the CS flux attachment to the fermion wavefunction, which leads to the $K=2$ CS theory at low energies. In 3D, we construct the state using a CS flux attachment on a 2D plane and a Jordan-Wigner transformation on a transversal vortex line with an arbitrary shape. We show that the derived action in (3+1)D space reproduces the Gauss law of this combined flux attachment with Jordan-Wigner transformation as a solution to the equations of motion. Furthermore, we compute the chemical potential of the CVLL state in the case of an infinitely long cylindrical moat band (when the dispersion has a minimum on an infinite cylinder in 3D reciprocal space) as a function of the particle density employing the Monte Carlo simulations. By further increasing the precision of the simulations in Ref.~\cite{wei2023chiral}, we show that in the dilute limit, the energy reproduces the equation of state, $\mu_{\text{cylinder}}\sim n^2\log ^2 (n)$, inherent to CSL. This result rules out the possibility of Bose condensation in such systems at low particle densities and suggests stabilizing the novel $U(1)$ symmetric CVLL ground state. 

The paper is organized as follows. Section II addresses the physics of interacting bosons, highlighting the role of the density of states in suppressing Bose condensation.  Emphasizing physical realizations, the section analyzes the diamond lattice structure as a prominent example, where frustration arising from competing nearest-neighbor and next-nearest-neighbor hopping integrals induces moat bands. Section III introduces the CVLL state as a ground state for interacting bosons in three-dimensional systems exhibiting moat-band dispersion. Section IV formulates the topological field theory underpinning the CVLL state. Section V develops a low-energy effective field theory describing interacting bosons with cylindrical moat band dispersion. 
Section VI investigates the CVLL ground state in interacting bosonic systems with cylindrical moat dispersion. We employ Monte Carlo simulations to quantitatively determine the chemical potential and equation of state, focusing explicitly on the dilute limit. 

The results of the present work have implications for the physics of frustrated quantum magnets, ultracold-atom systems, roton quasiparticles in  $^4$He, and the physics of moat regimes in heavy-ion collisions, which are discussed in the Outlook Section.

\section{The model}

In the present section, we will discuss the physics of the moat band for interacting bosons, emphasizing how the 1D-like divergence of the density of states near the bottom of the moat band leads to the absence of condensation. We will particularly explore the origins of moat bands in the contexts of frustrated quantum magnets and ultracold-atom systems. We analyze kinetic energy dispersions for some 3D systems, showing how these structures exhibit 1D-like behavior.

\subsection{Moat-band in 3D}

\begin{figure}[h]
\centering
\includegraphics[width=50mm,angle=0,clip]{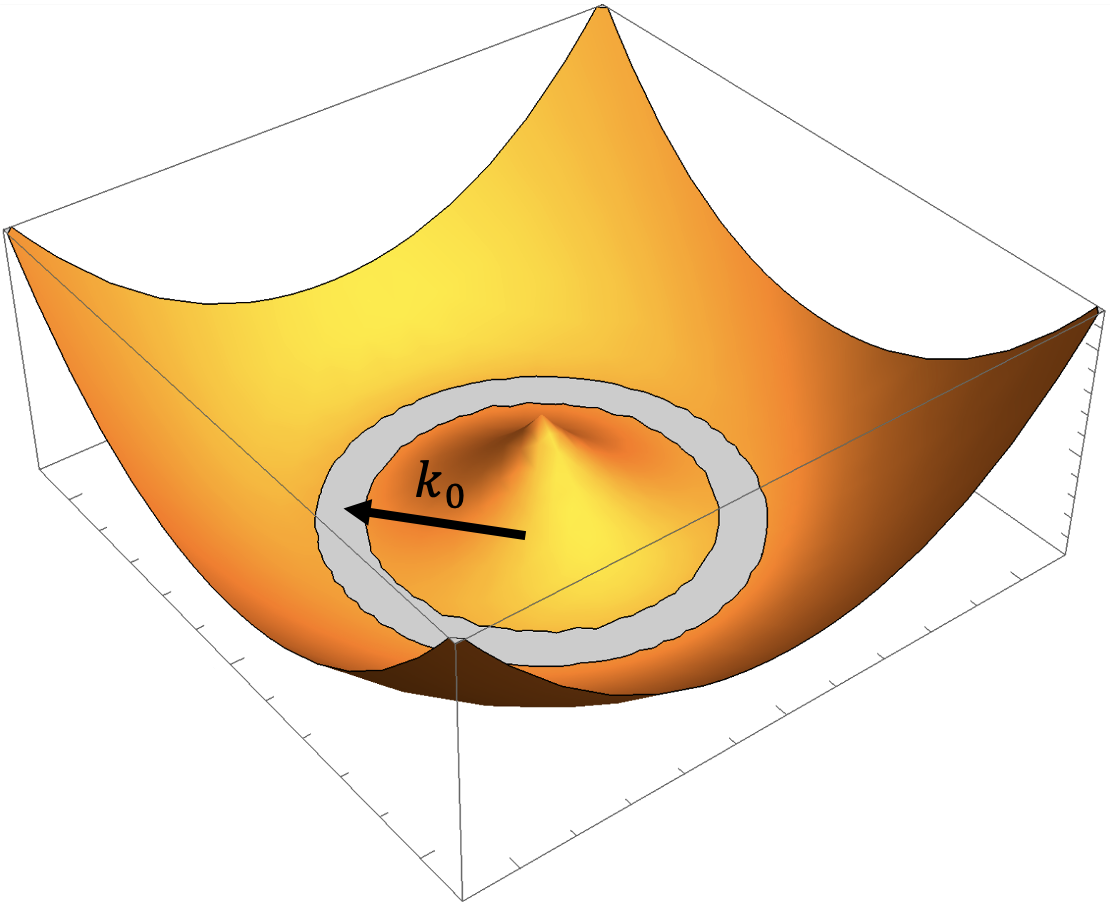}
\caption{(Color online) Moat band in 2D, illustrating a minimum energy contour forming a moat at a finite wavevector $k=k_0$. The band structure exhibits a local minimum along the closed manifold in momentum space, which is characteristic of systems with competing interactions with frustration. Such a moat dispersion for interacting bosons in 2D leads to forming the chiral spin (Bose) liquid supporting the semion topological order~\cite{sedrakyan-2012,sedrakyan-2015-1,sedrakyan-2015-2,sedrakyan-2023,wei2023chiral,sedrakyan-2014,emergent2022}.} 
\label{moatband}
\end{figure}

In the 2D \emph{moat band} scenario of single particle physics, where the degenerate energy minimum is distributed along a closed contour (see Fig. \ref{moatband}), the DOS in 2D behaves analogously to a 1D system~\cite{Capp-Gir-Mars-2007, Berg-2012, Chamon2014}. 
In lattice models, such dispersion is observed in a frustrated 2D honeycomb lattice~\cite{sedrakyan-2014}, square lattice with $\pi$ flux~\cite{Pi-square}, Floquet-driven square lattice~\cite{sedrakyan-2015-2, Floquet-Engineered-Moat}, and other bipartite lattices with frustration~\cite{sedrakyan-2014,sedrakyan-2015-1}. The low-energy physics and ground state of the interacting bosons in a moat band lead to the formation of the chiral spin liquid, stabilizing a semion topological order~\cite{sedrakyan-2012,sedrakyan-2015-1,sedrakyan-2015-2,sedrakyan-2023,wei2023chiral,sedrakyan-2014,emergent2022}. The interesting question is whether similar physics exists in 3D, where the degenerate minima would form surfaces instead of loops.

\begin{figure}[t]
\centering
\includegraphics[width=0.5\textwidth]{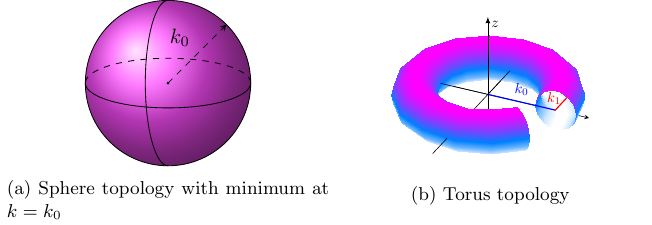}
\caption{Examples of 3D {\it moat} manifolds corresponding to dispersion relations (\ref{kinetic}).}
\label{3Dmanifolds}
 \end{figure}

Interestingly, the DOS still exhibits a 1D-like divergence near the bottom of a 3D moat band. We will show that this divergence profoundly impacts the system, suppressing Bose-Einstein condensation. While a 3D system with a parabolic dispersion would typically support condensation below a critical temperature, the 1D-like DOS in the moat band leads to enhanced fluctuations that destabilize the condensate, preventing long-range order.

To explore the low energy sector in 3D interacting bosonic moat-band systems, we define the kinetic energy dispersions for different manifold geometries as:
\bea
\label{kinetic}
&K_{\text{T}}({\bf k})=\frac{1}{2M}( \sqrt{(|k_{\text 2D}|-k_0)^2+k_z^2}-k_{1})^2,\nn\\
&K_{\text{S}}({\bf k})=\frac{1}{2M}(\sqrt{|k_{\text 2D}|^2+k_z^2}-k_{0})^2,\\
&K_{\text{C}}({\bf k})=\frac{1}{2M}(|k_{\text 2D}|-k_0)^2.\nn
\ena
Here, $k_{\text{2D}} = \sqrt{k_x^2 + k_y^2}$ and ${\bf k} = (k_x, k_y, k_z)$ are the 2D and 3D momenta, respectively. These equations describe the kinetic energy for toroidal ($K_{\text{T}}$), spherical ($K_{\text{S}}$), and cylindrical ($K_{\text{C}}$) geometries. In the torus geometry, $k_0$ is the major radius and $k_1$ the minor radius, with $k_1$ ensuring zero energy on the surface of the manifolds (see Fig. \ref{3Dmanifolds}). Regardless of the manifold, the DOS scales as $1 / \sqrt{\epsilon}$, characteristic of 1D systems, diverging as the energy $\epsilon$ approaches the lowest value in all 3D moat surface geometries.  This divergence at low energies amplifies quantum fluctuations that prevent the formation of a condensate. 
This behavior is intuitively similar to the absence of condensation in 1D systems, where the Mermin-Wagner theorem forbids the spontaneous breaking of continuous symmetries at finite temperatures. The moat band transfers this 1D behavior to higher-dimensional systems, suppressing condensation through enhanced fluctuations along the degenerate minima.

\subsection{Moat Lattice in 3D}

\begin{figure}[t]
\centerline{\includegraphics[width=50mm]{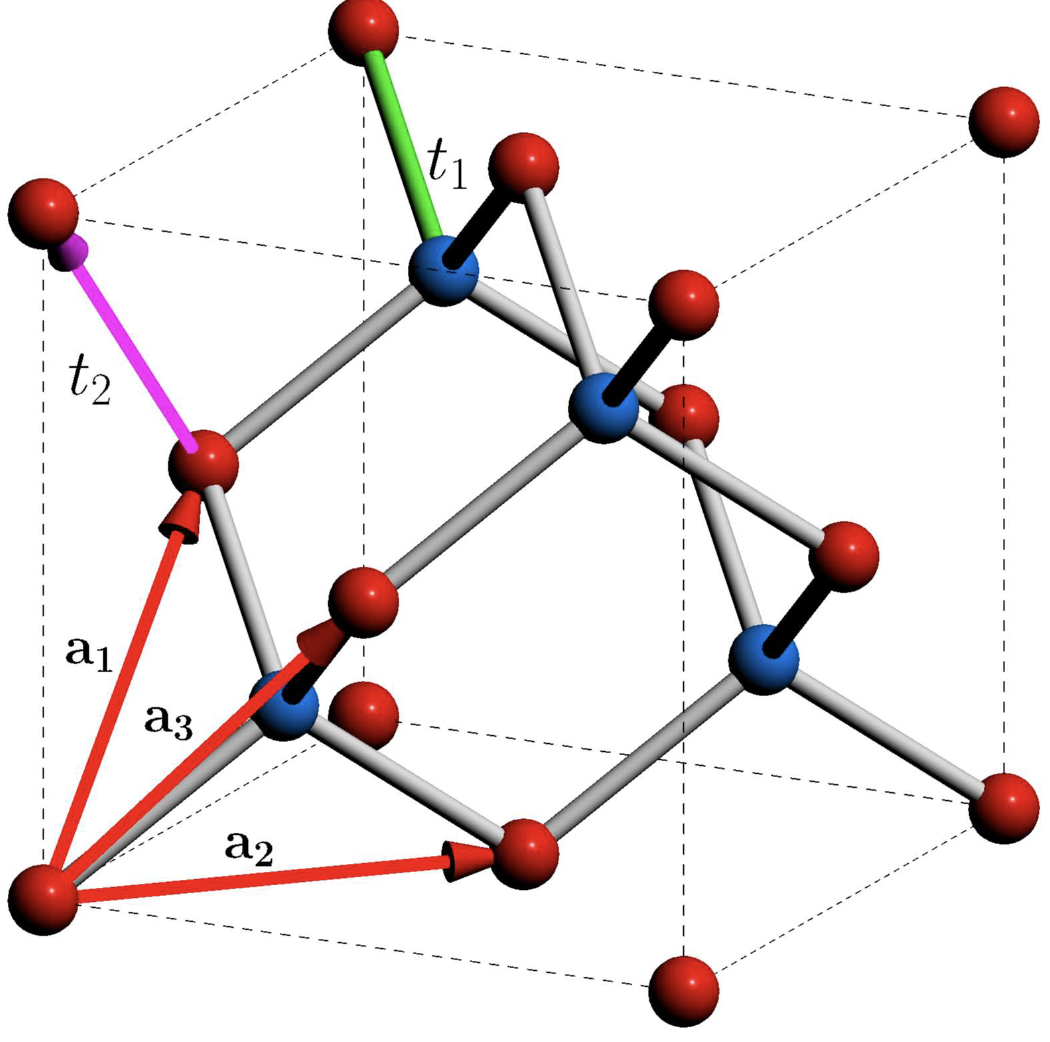}}
\caption{(Color online) 
A fragment of the Diamond lattice, illustrating the nearest-neighbor ($ t_1 $) and next-nearest-neighbor ($ t_2 $) hopping parameters. The lattice structure is defined by the primitive lattice vectors $ \mathbf{a_1} = \frac{a}{2}(0,1,1) $, $ \mathbf{a_2} = \frac{a}{2}(1,1,0) $, and $ \mathbf{a_3} = \frac{a}{2}(1,0,1) $.  The Diamond lattice, composed of two interpenetrating face-centered cubic sublattices, is a fundamental model with geometric frustration leading to a 3D moat dispersion in a wide parameter range.
} 
\label{diamondlattice}
\end{figure}


While the CSL state with semion topological order in two-dimensional moat-lattice antiferromagnets, such as the frustrated spin-$\frac{1}{2}$ XY model on the honeycomb lattice with nearest-neighbor and next-nearest-neighbor interactions has been reported in previous works \cite{sedrakyan-2015-1, sedrakyan-2015-2,sedrakyan-2014,emergent2022}, the natural question is whether one can have an extension of this physics in three dimensions. The frustrated diamond lattice naturally extends these ideas, supporting a 3D moat band. Here, we will discuss the moat band structure in the diamond lattice, a bipartite 3D analog of the honeycomb lattice, see Fig. \ref{diamondlattice}. A spin-$1/2$ quantum materials candidate for the diamond lattice spin-$1/2$ Heisenberg antiferromagnet is the A-site spinel materials with the typical chemical formula $\text{AB}_2\text{X}_4$, where the lattice has the next-nearest neighbor bond frustration \cite{LB-2008, Otimaa-2019, CMLSFI-2023, Bergman}. The existence of magnetism in such materials was experimentally observed in previous studies \cite{BLK-2008, faranak-2022, KYPSPD-2022}. Here, we propose the realization of the spin-$1/2$ XY antiferromagnet using ultracold atoms in the 3D optical diamond lattice, given the correspondence between the XY model and a system of hard-core bosons with strong on-site repulsion.

To this end, we focus on the Hamiltonian for hard-core interacting bosons on the diamond lattice. The lattice structure consists of two interpenetrating face-centered cubic (fcc) sublattices displaced by the vector $ a/4(1, 1, 1)$, where $ a $ is the diamond lattice constant. The Hamiltonian incorporates both nearest-neighbor (NN) and next-nearest-neighbor (NNN) antiferromagnetic exchange interactions, $t_1$ and $t_2$, correspondingly:
\bea
\label{Diamond1}
\hat{H}&=&t_1 \sum_{l=1}^{4} a^{\dagger}_i({\bf r})b_j({\bf r}+{\bm \delta}_{l}) + t_2  \sum_{k=1}^{12} a^{\dagger}_i({\bf r})a_j({\bf r}+{\bf e}_{k})\nn\\
&+&t_2  \sum_{k=1}^{12} b^{\dagger}_i({\bf r})b_j({\bf r}+{ {\bf e}_{k}}) +\mu \hat{n}(r),
\ena
where the summation over ${\bf r}$ representing the coordinates of each unit cell is assumed. Vectors ${\bm \delta}_{l}$, with $l=1\ldots 4$, and ${\bm e}_{k}$, with $k=1\ldots 12$, are the translation vectors between the nearest neighboring sites (NN) and the next nearest neighboring site (NNN), respectively.  Their explicit forms are $\bm{\delta}_{1}=\frac{a}{4}(1,1,1)$, $\bm{\delta_2}=\frac{a}{4}(1,-1,-1)$, $\bm{\delta_3}=\frac{a}{4}(-1,-1,1)$, and $\bm{\delta_4}=\frac{a}{4}(-1,1,-1)$ being the four translation vectors between NNs and $\bm{e_i}=\frac{a}{2}(\pm1,0,\pm1)$, $\bm{e_i}=\frac{a}{2}(\pm1,\pm1,0)$, and $\bm{e_i}=\frac{a}{2}(0,\pm1,\pm1)$ are the twelve translation vectors between NNNs. The bipartite nature of the diamond lattice allows for two annihilation operators for identical bosons, $a$ and $b$, which operate on different sublattices $A$ and $B$, correspondingly. 

In the $t_1$--$t_2$ Heisenberg antiferromagnet on the diamond lattice, the classical frustration threshold occurs at
$ \frac{t_2}{t_1} = \frac{1}{8}$ (this is when a degenerate spherical moat occurs in the single particle dispersion relation),
while quantum fluctuations shift the critical ratio to
$  \frac{t_2}{t_1} \simeq 0.18$~\cite{Otimaa-2019}. Motivated by the moat band physics in (3+1)D, we propose that in the XY interaction limit (hard core bosons), the conventional second-order transition at $  \frac{t_2}{t_1} \simeq 0.18$ from collinear Néel order to an incommensurate spiral is preempted by the emergence of a CVLL phase for intermediate frustration ratios $  \frac{t_2}{t_1} \in (0.125,0.18),$ and that a subsequent transition at
$  \frac{t_2}{t_1} \simeq 0.25$ marks the onset of spiral order from the CVLL state.

In Fourier space, the single-particle analog of Hamiltonian (\ref{Diamond1}) becomes:
\bea
\label{Diamond2}
\hat{H}=\sum_{k} \psi_{\bf k}^{\dagger} \left(\begin{array}{cc} t_2|F({\bf k})| & t_1f({\bf k}) \\  t_1f^*({\bf k}) & t_2|F({\bf k})|  \end{array}\right) \psi_{\bf k},
\ena
where $\psi_{\bf k}^{\dagger}=(a^{\dagger}_{\bf k},~b^{\dagger}_{\bf k})$, 
\begin{eqnarray}
\!\!\!f({\bf k})=4\left(\cos \frac{k_xa}{4} \cos \frac{k_ya}{4} \cos \frac{k_za}{4}\right. \nonumber\\
\left.\qquad -i \sin \frac{k_xa}{4}\sin \frac{k_ya}{4} \sin \frac{k_za}{4} \right), 
\end{eqnarray}
and $F({\bf k})=|f({\bf k})|^2-4$. One can write the Hamiltonian in a simpler form by defining
\begin{eqnarray}
\hat{T}_{\bf k}=\left(\begin{array}{cc} 0 & f({\bf k}) \\  f^*({\bf k}) & 0 \end{array}\right). 
\end{eqnarray}
Therefore, one will have for the Hamiltonian Eq.~(\ref{Diamond2}): $\hat{H}=t_1\hat{T}_{\bf k}+t_2(|\hat{T}_{\bf k}|^2-4\hat{I})$, where $\hat{I}$ is the 2D identity matrix. It is straightforward to find the two energy branches as follows
\bea
E^{(\pm)}({\bf k})=\pm t_1 |f({\bf k})|+t_2(|f({\bf k})|^2-4).
\ena
\begin{figure}[!t]
  \centering
{\includegraphics[width=0.23\textwidth]{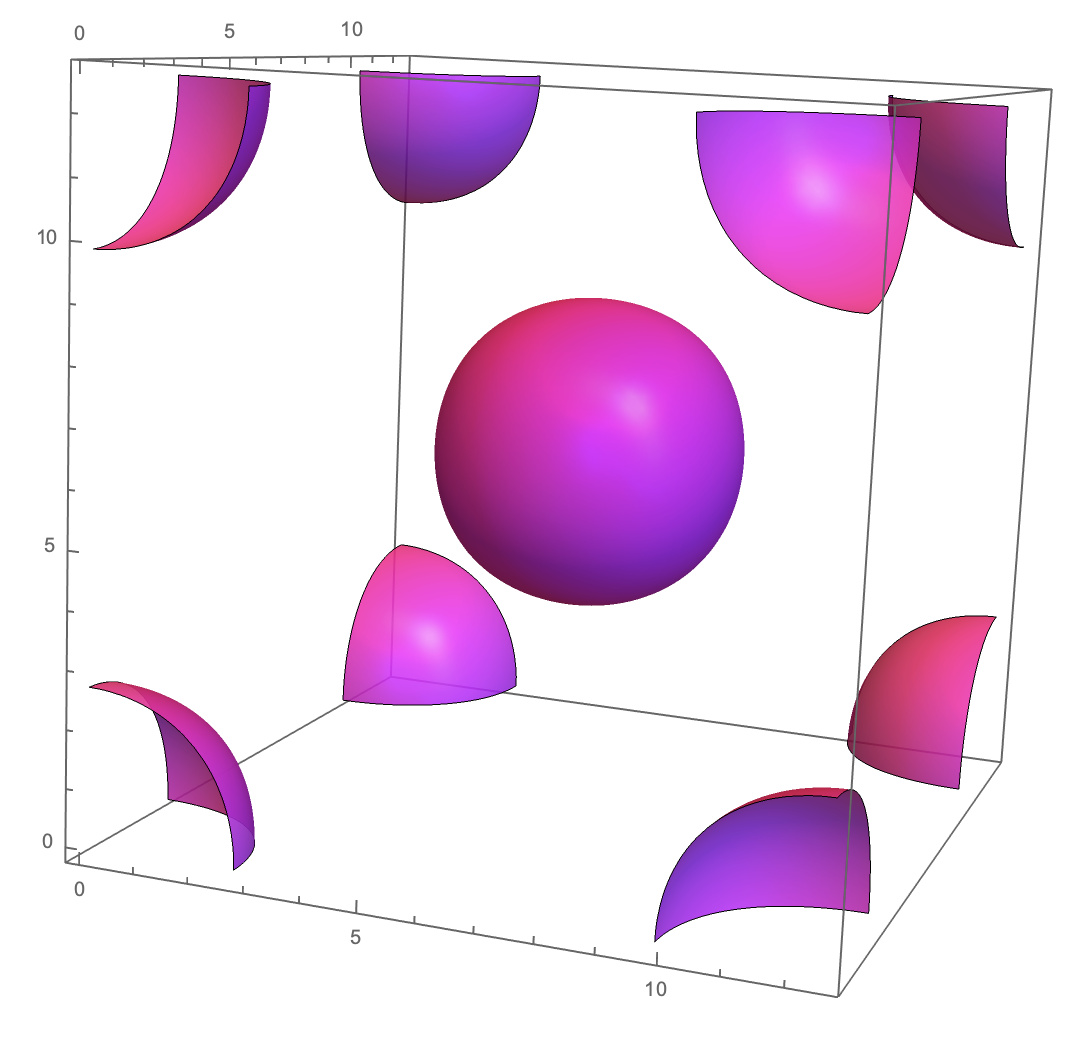}\label{spherical}}
  \hfill
{\includegraphics[width=0.23\textwidth]{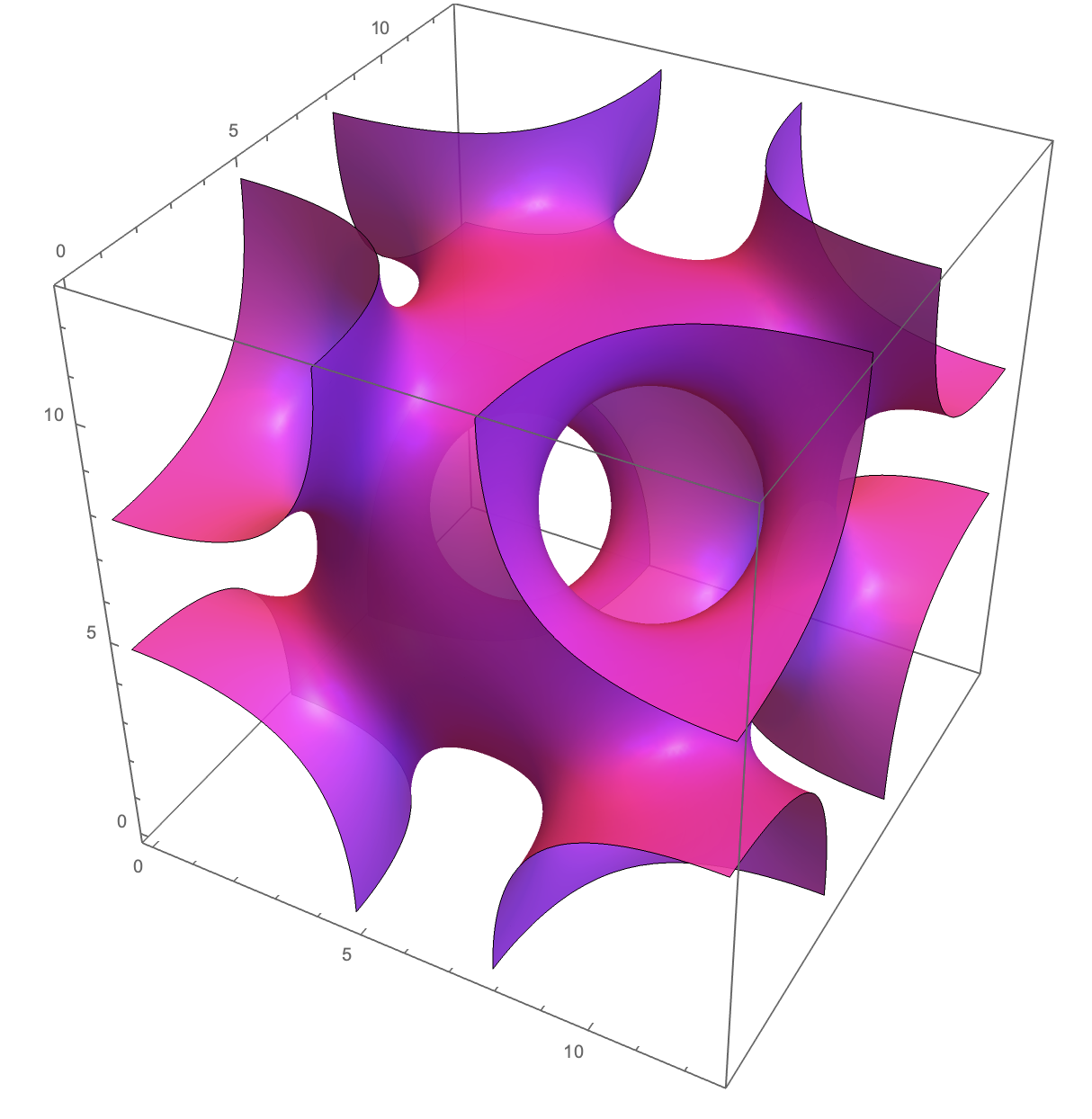}\label{holes}}
  \caption{(Color online) Infinitely degenerate minimum equi-energy surface, the 3D moat, corresponding to the minimum of the single-particle dispersion relation on a three-dimensional Diamond lattice for different ratios of $ t_2/t_1 $. The surface represents the continuum loci of momentum-space points where the energy reaches its minimum, highlighting the role of the next-nearest-neighbor hopping parameter $ t_2 $ in modifying the band structure. As $ t_2/t_1 $ varies, the topology and shape of the degenerate manifold evolve, indicating the presence of frustration-driven instabilities. Left panel: The moat surfaces around the $\Gamma$ point of the Brillouin zone at $1/8<t_2 /t_1<1/4$. For $t_2 /t_1<1/8$, the dispersion has a single minimum at the $\Gamma$ point. Right panel: The moat surfaces touch at $t_2 /t_1=1/4$, and merge as $t_2 /t_1>1/4$.
  }\label{minimum}
 \end{figure}
Since $t_1$ and $t_2$ are positive, the lowest energy branch is $E^{(-)}({\bf k})=- t_1 |f({\bf k})|+t_2(|f({\bf k})|^2-4)$. The minimization of this energy with respect to the momentum, $k$, yields the following equation $|f({\bf k})|=\frac{t_1}{2t_2}$. 
This is an equation for the 3D wavevector, ${\bf k}$, in a 3D reciprocal space. In general, the solution to this equation defines a 2D manifold embedded in the 3D space — the moat.

When $\frac{t_2}{t_1}\leq\frac{1}{8}$, the equation of minimum dispersion has a solution in the form of a single minimum at the $\Gamma$ point of the Brillouin zone. This suggests that the bosons for such values of the exchange integrals will condense at the $\Gamma$ point, consistent with the previously obtained collinear Néel phase for small $t_2$. At $\frac{t_2}{t_1}=\frac{1}{8}$, initially the surface solution for the equation of minimum dispersion starts to form a sphere (when the ratio $t_2/t_1$ is slightly greater than $1/8$), and as the ratio increases, the formation of the surface changes according to the symmetries of the lattice.  The minimal contour is topologically a sphere for $1/8< t_2/t_1< 1/4$ centered at the $\Gamma$ point. At $t_2/t_1=1/4$, the minimal surface becomes multiply connected, and at $t_2/t_1>1/4$, the surfaces merge into a larger multiply connected surface with holes.
The evolution of the minimum surface is shown in Fig. \ref{minimum}. 
The situation here is similar to the formation of the moat contour in the frustrated 2D honeycomb lattice~\cite{sedrakyan-2014}.
The formation of the moat surface for bosons at $\frac{t_2}{t_1}=\frac{1}{8}$, classically, signals a transition from the condensate at the $\Gamma$ point to the condensate at finite momentum, corresponding to an incommensurate spiral state~\cite{Bergman}. 

In the case of hard-core bosons, strong on-site repulsion prevents double occupancy of any site. This constraint allows us to map the hard-core boson system onto an equivalent quantum magnet. Specifically, the bosonic creation and annihilation operators on sites positioned at ${\bf r}_A$ and ${\bf r}_B$ of sublattices $A$ and $B$, correspondingly, can be equivalently redefined as spin-$\frac{1}{2}$ operators:
\begin{eqnarray}
a^\dagger({\bf r}_A) \to {S}_{{\bf r}_A}^+, \quad a({{\bf r}_A}) \to S_{{\bf r}_A}^-,\nonumber\\
b^\dagger({\bf r}_B) \to {S}_{{\bf r}_B}^+, \quad b({\bf r}_B) \to S_{{\bf r}_B}^-,
\end{eqnarray}
where $S_{\bf r}^+$ and $S_{\bf r}^-$ are the spin raising and lowering operators, respectively. These operators satisfy the commutation relations for spin-$\frac{1}{2}$ particles and enforce the hard-core condition by limiting each site to either zero or one boson. 
The Hamiltonian for hard-core bosons with NN hopping and strong on-site repulsion takes the form equivalent to the XY antiferromagnetic model:
\begin{align}
&\hat{H}_{\text{XY}} = \\\nn 
&t_1 \sum_{\langle i,j \rangle} (S_i^{+} S_j^{-} + h.c)+t_2 \sum_{\langle\langle k,l \rangle\rangle} (S_k^{+} S_l^{-} + h.c) + h\sum_i S_i^z,
\end{align}
where $h$ is the external magnetic field in the direction of the $z$-component of the spins, $\langle \ldots \rangle$ corresponds to the summation performed over the NN bonds while 
$\langle\langle \ldots \rangle\rangle$ corresponds to the summation performed over the NN bonds on the 3D diamond lattice. 
This equivalence provides a framework for analyzing strongly interacting hard-core bosons on the diamond lattice, leveraging a well-developed tool in quantum magnetism.
The previous investigation of the frustrated Heisenberg antiferromagnet on the diamond lattice suggested a second-order phase transition from a collinear Néel phase to an incommensurate spiral phase at $t_2/t_1 \approx 0.18$, highlighting the role of quantum fluctuations in stabilizing novel phases beyond the classical limit of $t_2/t_1 = 1/8$. In the language of hard-core bosons, the fluctuations shift the transition from the condensate at $\Gamma$ point to an incommensurate spiral with condensate wave vector $(\tilde{k}, \tilde{k}, 0)$ at $t_2/t_1 \sim 0.18$, which is considerably higher than the classical value $1/8$~\cite{Bergman, LB-2008, Otimaa-2019}. 
It was also shown that in systems with tetragonal distortion, such as CuRh$_2$O$_4$, spiral phases become energetically favorable along specific crystallographic directions.

Our main results below for the continuum limit of a bosonic moat-band system in 3D lead us to conjecture that the diamond lattice spin-1/2 $XY$ antiferromagnet near full magnetization in $z$ direction stabilizes the CVLL ground state at sufficiently strong frustration of intermediate values of $t_2/t_1$ between the collinear Néel phase and the incommensurate spiral phase when the moat band effect is the strongest. 

In conventional Bose systems, condensation corresponds to the macroscopic occupation of a single particle state, accompanied by spontaneous $U(1)$ symmetry breaking. However, the frustrated system prevents such condensation in the moat band scenario.
A key implication of our present study in the diamond lattice spin-1/2 XY antiferromagnet is the absence of ordering and stabilization of the time-reversal symmetry breaking CVLL phase between the condensate at the $\Gamma$ point and the condensate at finite momentum on the moat. This, in turn, implies stabilization of chiral order parameters~\cite{sedrakyan-2015-1, Maiti-2019,TS-TJ,TJ2,emergent2022,wei-2024-1,wei-2024-2,Poincare,Daniel,Daniel2,lake,sheng} on triangles composed by the NN and NNN links in the unit cell of the 3D diamond lattice. In the following subsection, we discuss the equations of state for the condensate states in a moat band. We will subsequently calculate the equation of state for the CVLL and demonstrate that it has a lower energy compared to the condensate energies.

\subsection{Condensate on a spherical moat band}

In general, the energy of a Bose condensate reflects the interplay between the kinetic energy and the interaction energy of the constituent particles. For weakly interacting dilute gases, the condensate energy per particle can be expressed in terms of the particle density $n$ and the interaction strength $g_0$, as
$E_{\text{BEC}} \sim g_0 n$. This relationship in dilute Bose gases arises because the interactions are effectively captured by a single parameter, the scattering length, which defines the bare coupling strength $g_0$. The renormalization of the coupling constant, $g_0\rightarrow g_{\text{univ}}$, with $g_{\text{univ}}$, accounts for short-range correlations and ensures that the theory accurately describes the low-energy behavior of the system, even though the microscopic interactions may be more complex. The proportionality of the condensate energy to $n$ reflects the mean-field nature of the interactions, where each particle experiences the average effect of all other particles in the system.
This is the universal behavior of the energy, which is valid for all $U(1)$ symmetry-breaking BECs at low densities.

In examining the energy of a Bose condensate consisting of interacting bosons within a 3D spherical moat band as an example, it is instructive to consider several condensate types and identify which condensate state minimizes the interaction energy, yielding a more stable configuration. We are interested in dilute Bose gases with the s-wave short-range interaction Hamiltonian given by
\begin{equation}
H_{\text{int}}=\frac{g_0}{2M}\sum_{i,j}^{N} \delta({\bf r}_i - {\bf r}_j),    
\end{equation}
where $M$ is the particle mass and $N$ is the total number of particles in the system. 

Let us consider the following two ways to construct the condensate wave function for this many-body system, each associated with a different energy outcome. First, consider a Bose condensate composed of identical single-particle states, with all particles in the same momentum state with  $|\mathbf{k}| = k_0$.
The corresponding many-body wave function can be expressed as
\bea
\Psi_B^{(0)}({\bf r})=\prod_{i=1}^{N} e^{i \mathbf{k} \mathbf{r}_i},
\ena
where $e^{i \mathbf{k} \mathbf{r}_i}$ represents the plane-wave state for the $i$-th particle. This wave function describes a condensation with all particles in the same momentum state, forming a coherent, uniform condensation. The interaction energy for this configuration is determined by the expectation value of the interaction Hamiltonian $H_{\text{int}}$. Calculating the energy per particle for this state yields
\begin{equation}
E^{0}_{\text{int}}=\frac{g_0 N^2}{4MV},
\end{equation}
where $V$ is the system volume. 

Secondly, consider a condensate formed by a superposition of two distinct momentum states, $ \mathbf{k}_1$ and $ \mathbf{k}_2$, both lying on the same degenerate manifold such that $ |\mathbf{k}_1| = |\mathbf{k}_2| = k_0 $. The wave function for this mixed-state condensate can be written as
\bea
\Psi_B^{\phi}({\bf r})=\prod_{j=1}^{N}\frac{1}{\sqrt{2}} [e^{i \mathbf{k}_1 \mathbf{r}_j}+e^{i \mathbf{k}_2 \mathbf{r}_j}].
\ena
This superposition introduces interference between the two momentum components, leading to a more complex condensate structure than the uniform plane-wave configuration. The energy corresponding to this mixed-state condensate is 
\begin{equation}
E^{\phi}_{\text{int}}=\frac{3g_0 N^2}{4MV}.    
\end{equation}

A comparison of the two energies reveals that
$E^{0}_{\text{int}} < E^{\phi}_{\text{int}},$
indicating that the uniform condensate state $ \Psi_B^{(0)} $ has a lower interaction energy than the superposed state $ \Psi_B^{\phi}$. This outcome aligns with the expectation that a more coherent condensate, without interference between distinct momentum components, minimizes the interaction energy, making it energetically more favorable.

More generally, the moat dispersion in 3D interacting bosonic systems and fluctuations can drive the homogeneous condensation of the system towards forming inhomogeneous phases~\cite{Braz-1975}. This results in spatially modulated condensates, where the system may form phase-modulated states or amplitude-modulated structures corresponding to a one-dimensional modulation embedded in 3D space. Both configurations, driven by the finite-momentum minimum, avoid the homogeneous condensates. 
The fluctuation-induced interactions can alter the nature of phase transitions in systems with spatially modulated order parameters. In particular, the effect of fluctuations involves the thermal (Brazovskiǐ) and quantum (Dyugaev) fluctuations~\cite{Braz-1975, Dyug-1975} that introduce a first-order phase transition instead of the second-order transition. This is attributed to the singular behavior of the fluctuation field’s propagator near the transition point, which manifests as a divergence at $k=k_0$ due to the loop integrals of the fluctuating fields.

The effect is notable because fluctuations modify both second- and fourth-order vertices in the thermodynamic potential. Specifically, the fourth-order vertex’s fluctuation-induced modification changes sign, enforcing a first-order transition. This result aligns with Brazovskiǐ's findings that thermal fluctuations adjust the coefficients of higher-order terms, driving a shift from second-order to first-order transitions. Dyugaev’s work extended this phenomenon to quantum fluctuations, showing similar effects in quantum systems at zero temperature, where these fluctuations influence the vertex functions even in the absence of thermal contributions.

To derive the scaling of the chemical potential, $\mu$, with the condensate order parameter in a spatially modulated phase at zero temperature, we consider the effective action for a system where the bosonic condensate is characterized by a nonzero wave vector $k_0$ due to the moat dispersion. Following the Brazovskiǐ theory framework, the thermodynamic potential $\Omega$ includes contributions up to the fourth order in the order parameter $ \psi $ (assumed real for simplicity) and is given as:
\begin{equation}
    \Omega(\psi) = \frac{1}{2} \tau \psi^2 + \frac{1}{4} \bar{\lambda} \psi^4 + \cdots,
\end{equation}
where $ \tau = \mu - \mu_c $, with $ \mu_c $ being the critical chemical potential at which the transition to the inhomogeneous phase occurs. The parameter $\bar{\lambda}$ is the renormalized fourth-order coupling constant. Minimizing $ \Omega $ with respect to $ \psi $, one finds the equilibrium condition for the condensate, yielding the scaling relation $ \psi \sim \sqrt{\mu - \mu_c}$. This quadratic dependence between $\mu $ and $ \psi $ arises from the $\psi^4$-term in the thermodynamic potential and implies that the chemical potential increases quadratically with the order parameter.

For a dilute, inhomogeneous condensate, the particle density behaves as $n \sim \psi^2 $ since $n$ measures the number of condensed particles per unit volume. Substituting $ \psi^2 $ for $ n $ in the chemical potential scaling, we obtain the linear relationship: $\mu \sim g_{\text{univ}} n,$ where $ g_{\text{univ}} = \bar{\lambda} / 2 $ is the renormalized coupling constant. 

Below, we will argue that the 3D extension of the CSL, namely the proposed CVLL state, is energetically more favorable at low densities compared to the spatially modulated condensate states discussed above. In a CVLL, topological defects such as vortex lines become the primary degrees of freedom, leading to a phase in which the system can avoid the costs associated with long-range coherence that is typical of conventional condensate states. This is in contrast to the inhomogeneous condensate phases where the moat dispersion forces bosons to condense at finite momentum, demanding $\sim g_{\text{univ}} n$ scaling of chemical potential with density.

\section{The chiral vortex-line liquid state} 

The presence of a moat dispersion, with its characteristic minimum at finite momentum, introduces a one-dimensional-like density of states that enhances the likelihood of a lower energy of the fermionized wavefunction without a conventional condensate fraction. This structure suggests that for a many-body bosonic system, a non-condensed state, such as the CSL in two dimensions, could offer a lower chemical potential compared to spatially modulated condensate phases. Building on this intuition, we seek to construct a quantum state for N bosons that fulfills two essential criteria: (1) the wavefunction must vanish when particle coordinates coincide, mimicking the anti-symmetric properties of a fermionic Slater determinant and thereby avoiding bosonic clustering; and (2) a mechanism, similar to a CS field, should introduce phase factors that effectively transmute the bosonic statistics, precluding macroscopic occupation of any single state and thus ensuring the absence of a condensate fraction. This approach aims to establish a ground state that employs fermionic exclusion principles within a bosonic system, leveraging the one-dimensional density of states induced by the moat dispersion to energetically favor such non-condensed, topologically ordered phases.

In 2D, the CS flux attachment allows the representation of an $N$-body bosonic wavefunction in terms of an antisymmetric one by attaching phase-dependent fluxes to each particle in a two-dimensional system. This transformation introduces a complex phase factor to the bosonic wavefunction, multiplied by the antisymmetric $N$-body wavefunction, an inherent property of fermionic states. Specifically, the CS transformation reads:
\bea
\label{CS-1}
\Psi_B(w_1,\cdots w_N)=\prod_{i>j} \left(\frac{w_i-w_j}{|w_i-w_j|}\right)^m \:\Psi_F (w_1,\cdots w_N)\nonumber\\
\ena 
where $w_j=x_j+i y_j$ with $j=1,..., N$ represent the planar coordinates of particles in complex form, and $m=1,3,5\ldots $ denotes the flux number, an odd integer, signifying the strength of the attached flux. The fermionic wavefunction, $\Psi_F(w_1, \cdots, w_N)$, when multiplied by this CS phase factor, becomes symmetric with respect to particle exchange, effectively transforming it into a bosonic wavefunction $\Psi_B(w_1, \cdots, w_N)$. 

This CS flux attachment transformation in 2D can technically apply to any wavefunction, including a Bosonic wavefunction representing a condensate. However, in the case of a condensate, the CS flux attachment becomes trivial. The reason is that in a standard condensate, the antisymmetric part of the wavefunction would possess a conjugate phase that exactly cancels the phase contribution from the CS factor. This results in no net effect of the flux attachment, as if each boson carries a flux opposite to that of its neighbors, leading to mutual cancellation. An example of such a ``fermionized" state, where the CS phases are canceled, is provided by the work of Girvin and MacDonald~\cite{gir-mac}, which demonstrated that the absolute value of the Slater determinant wavefunction—without the net CS phase factor—has a finite condensate fraction, representing a symmetry-broken condensate phase. In this phase, Goldstone modes arise as gapless excitations, indicating the presence of long-range order rather than topological order. This is in contrast to a topologically ordered phase, where gapless Goldstone modes would be absent.

To ensure that a bosonic many-body wavefunction achieves true statistical transmutation and topological ordering, it is essential that the fermionic part of the wavefunction, $\Psi_F(w_1, \cdots, w_N)$, carries a phase structure that matches—not cancels—the “statistical” CS phase factor. This alignment ensures that the CS phase factor does not negate itself upon particle exchange. Instead, it introduces a robust topological structure to the bosonic system, effectively transforming its statistics and enforcing the absence of macroscopic occupation, typical of a non-condensed, topologically ordered phase. This careful alignment of phases is thus crucial to creating a fermionized bosonic system where topological order replaces conventional Bose-Einstein condensation, with particles avoiding the occupation of a single quantum state in favor of a collective, fractionalized behavior.

The CS flux attachment can be described in the second quantized formalism, where we express the bosonic field operator $ \hat\Psi_B $ in terms of the fermionic field operator $ \hat\Psi_F $. In 2D space, the second quantized form of the CS transformation is given by~\cite{Halperin-1993}
\bea
\label{CS-2}
\hat\Psi_B^{2D}({\bf r}_i)= \exp\Big\{i m \sum_{j\neq i}\arg({\bf r}_i-{\bf r}_j)\hat n_j \Big\}\hat\Psi_F^{2D} ({\bf r}_i).
\ena 
The function $ \arg({\bf r}_i - {\bf r}_j)$ represents the angle between the 2D radius vectors of particles $i$ and $j$, while $\hat n_j = (\hat\Psi_F^{2D})^\dagger \hat\Psi_F^{2D} = (\hat\Psi_B^{2D})^\dagger \hat\Psi_B^{2D} $ denotes the particle density operator. 

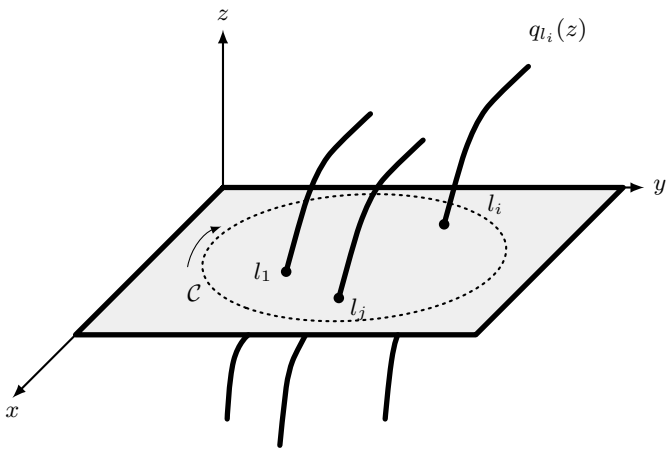
\begin{figure}[t]
\begin{tikzpicture}[line cap=round,line join=round, x={(-0.4cm,-0.4cm)}, y={(0.8cm,0)},z={(0,0.6cm)}, scale=0.7]

 	\coordinate (x1) at (0,0,0);
	\coordinate (x2) at (7,0,0);
	\coordinate (x3) at (7,9.5,0);
	\coordinate (x4) at (0,9.5,0);
\draw[axis] (0,0,0) -- (10,0,0) node[below] {$x$};
\draw[axis] (0,0,0) -- (0,10,0) node[right] {$y$};
\draw[axis] (0,0,0) -- (0,0,5) node[above] {$z$};

\draw[draw=black, fill=gray!50!white, nearly transparent,thick] (x1) -- (x2) -- (x3)-- (x4); 
\draw[draw=black, line width=2pt] (x1) -- (x2) -- (x3)-- (x4)--(x1); 
\draw [black, line width=2pt] plot [smooth, tension=0.5] coordinates { (1,2,-2)(2,3,1)(3,4,3)(4,5.5,5)}; 
\draw [black, line width=2pt, xshift=3cm, yshift=0.9cm] plot [smooth, tension=0.5] coordinates { (1,2,-2)(2,3,1)(3,4,3)(4,5.5,5)}; 
\draw [black, line width=2pt, xshift=1cm, yshift=-0.5cm] plot [smooth, tension=0.5] coordinates { (1,2,-2)(2,3,1)(3,4,3)(4,5.5,5)}; 
\draw [black, line width=2pt] plot [smooth, tension=0.8] coordinates { (7,4.1,0)(7.5,4,-0.5)(8,4.1,-2)};
\draw [black, line width=2pt, xshift=3cm]plot [smooth, tension=0.9] coordinates { (7,3.9,0)(7.5,4,-0.5)(8,4.1,-2)};
\draw [black, line width=2pt, xshift=1cm, yshift=-0.5cm] plot [smooth, tension=0.5] coordinates {  (7,4.2,0.8)(7,3.9,0)(7.5,4,-0.5)(8,4.1,-2)};
\fill (1,2,-2) circle[radius=1mm] ;
\fill (1,2,-2) circle[radius=1mm, xshift=3cm, yshift=0.9cm] ;
\fill (1,2,-2) circle[radius=1mm, xshift=1cm, yshift=-0.5cm] ;
 \draw[]  (1,2,-2) node[left,xshift=-2pt] {$l_1$};
 \draw[]  (1,2,-2) node[left,xshift=3cm, yshift=0.9cm] {$l_i$};
  \draw[]  (1,2,-2) node[left,xshift=1.2cm, yshift=-0.5cm] {$l_j$};
  \draw[]  (4,5.5,5.5) node[left,xshift=3cm, yshift=0.9cm] {$q_{l_i}(z)$};
   \draw[rotate=-20, thick, dotted] (1,4,0) ellipse (4 and 2.7) ;
 \draw[-latex,rotate=-10] ( (4,1.5,0)  arc (180:120:1cm);
 
 \draw (5,2.2,0) node[left]{$\mathcal{C}$} ;

  \end{tikzpicture}
  \caption{Real-space schematic of a segment of the 3D system, illustrating the structure emerging after representing the many-body bosonic wave function as a Slater determinant with effectively 2D CS phase factors in the XY plane and monotonic vortex lines along which Jordan-Wigner phases are introduced. Each vortex line is characterized by the function $ q^{}_{{\bf l}^{}_i}(z) $ and originates at the base point $ {\bf l}_i = (x_i, y_j) $, where it crosses the XY plane. The depicted 3D CS transformation (Eq.~(\ref{CS-3})) represents an efficient variational ansatz for the ground state of the many-body Hamiltonian, capturing the interplay between topological phases and many-body correlations and providing insight into the emergent quasiparticle statistics in this strongly interacting system. 
  \label{vortex-lines}}
\end{figure}

The above definition of the CS transformation is designed for 2D systems, where the phase factor accounts for particle exchanges in the plane. However, a natural question arises: How can this transformation be extended to 3D, where particles can exchange along the third spatial coordinate? To extend the CS transformation into 3D space, we must generalize the phase factor to respect the three-dimensional geometry. To this end, we introduce a function $q^{}_{\mathbf{l}}(z)$ that is designed to vary continuously with the third spatial coordinate $z$ and defines a mapping from a plane (at $z = 0$) to 3D space. This function is written as
 \bea
  \label{APtransformation}
   q\strut_{\bf l}(z)=x f(z) + i y/ f(z), \;\;\; f(0)=1,
\ena   
where ${\bf r} = ({\bf l}, z)$, with ${\bf l} = (x, y)$, representing the position of a particle in 3D space, and $f(z)$ is a monotonic function of $z$, which we will keep as a variational variable and pick the one which minimizes the energy most efficiently for the given type of the moat band. Upon fixing the coordinates ${\bf l} = (x, y)$ on the plane $z=0$, the function $ q^{}_{\bf l}(z)$ defines a continuous trajectory in 3D by varying $z$ - {\it the vortex line}. Points on this vortex line are parameterized by $(x f(z), y / f(z), z)$, where $x$ and $y$ remain constant as $z$ varies, as shown in Fig.~\ref{vortex-lines}. This transformation 
$(x,y,0)\rightarrow (x'=x f(z), y'=y/f(z), z)$ is an area preserving transformation of planes, since $dS = dxdy = dx'dy'$, ensuring that particle exchanges in 3D can be consistently represented.
The monotonicity of function $f(z)$ ensures that the number of points at which vortex lines cross at any higher or lower plane with fixed $z=\text{const}$ is the same, $L$.
Thus, each vortex line can be marked by the index ${\bf l}_i=(x_i,y_i),\; i=1,..., L $ representing the crossing point of the vortex line parametrized by $q^{}_{{\bf l}_i} (z)$
with $z=0$ plane, as in Fig.~\ref{vortex-lines}.

In the second quantized form, we extend the CS transformation to 3D as follows:
\bea
\label{CS-3} 
\hat\Psi_B({\bf r}_i)&=&\exp\Big\{i m \sum_{\substack {{\bf l}^\prime\neq {\bf l},z^\prime}}
\arg\big[q^{}_{\bf l}(z)-q^{}_{\bf l^\prime}(z^\prime)\big]\hat n_{\bf r^\prime} \Big\}\nn\\
&\times& \exp\Big\{i \pi \sum_{\substack{{\bf l^\prime}={\bf l}\\
z^\prime < z}}
\hat n_{{\bf r}^\prime}\Big\}\hat\Psi_F({\bf r}_i),
\ena
The first exponential term is an extension of the 2D CS phase factor to 3D, where the arguments $ q^{}_{\bf l}(z) $ and $ q^{}_{\bf l'}(z') $ introduce position-dependent phases based on particle positions along the $z$-axis. The second exponential factor acts as a ``string operator" that extends into the $z$-direction, representing a 1D Jordan-Wigner transformation along the vortex line. This operator introduces an additional phase factor when particles are ordered along the vortex line, ensuring that exchanges in the third dimension contribute to the overall phase consistently with fermionic antisymmetry.

The 3D CS transformation in the second quantization leads to a specific ansatz for the first quantized form of the many-body bosonic wavefunction. This wavefunction can be written as
\bea
\label{wave-function}
\Psi_B({\bf r}_1,\cdots {\bf r}_L)&=&\prod_{{\bf l}_i\neq {\bf l}_j}\left(\frac{q^{}_{{\bf l}_i}(z_i)-q^{}_{{\bf l}_j}(z_j)}{\mid q^{}_{{\bf l}_i}(z_i)-q^{}_{{\bf l}_j}(z_j) \mid}\right)^m\\
&\times&\prod_{{\bf l}_{i'} = {\bf l}_{j'}}\text{sign}[z_{i'}-z_{j'}] \Psi_F({\bf r}_1,\cdots {\bf r}_L)\nn
\ena
where $\Psi_F({\bf r}_1, \cdots, {\bf r}_L)$ is defined by a Slater determinant of single-particle wavefunctions $\psi_i({\bf l}_i, z_i)$ for $ i = 1, \cdots, L $.
In this wavefunction, the product over $ {\bf l}_i \neq {\bf l}_j $ incorporates the CS phase factors from the positions $ q^{}_{{\bf l}_i}(z) $ and $ q^{}_{{\bf l}_j}(z) $, ensuring that each pairwise exchange in plane at height $z$ includes the proper phase shift. The product over $ {\bf l}_{i'} = {\bf l}_{j'} $ introduces a sign factor depending on the relative $ z $-coordinates, which is crucial for enforcing antisymmetry in the third dimension. This overall structure results in a bosonic wavefunction that acquires a phase structure and a fermionic Slater determinant wavefunction, guaranteeing statistical transmutation and topological order in the 3D setting.

Let us examine the terms in Eq.~(\ref{CS-3}) in greater detail to understand their roles in the transformation process for mapping bosonic to fermionic operators in 3D space.
Consider the first term in the phase factor:
\begin{equation}
U_1({\bf r}) = \exp\left\{i m \sum_{{\bf l^\prime}\neq {\bf l}, z^\prime}\text{arg}\left[q^{}_{\bf l}(z) - q^{}_{\bf l^\prime}(z^\prime)\right] n_{\bf r^\prime} \right\}.
\end{equation}
This term corresponds to the 2D CS transformation, where the relative azimuthal angle between particles on the plane sets the phase factor. Here, $U_1$ accumulates a phase from the azimuthal angle between particle positions at different points ${\bf r}$ and ${\bf r}^\prime$ in the plane. Importantly, $U_1$ excludes points on the same vortex line, $q^{}_{\bf l}(z)$, to avoid phase singularities that could arise if ${\bf r}$ and ${\bf r}^\prime$ coincided. This phase term ensures that commuting bosonic operators are transformed into fermions, replicating the effect seen in the original 2D CS framework.

The second phase term in Eq.~(\ref{CS-3}) introduces a phase along the $z$ axis:
\begin{equation}
U_2({\bf r}) = \exp\left\{i \pi \sum_{{\bf l^\prime} = {\bf l}, z^\prime < z} n_{{\bf r}^\prime}\right\}.    
\end{equation}
This phase is non-zero when the points ${\bf r}$ and ${\bf r}^\prime$ lie on the same vortex line, meaning ${\bf l}^\prime = {\bf l}$, but at different $z$ coordinates, $z \neq z^\prime$. It functions similarly to the Jordan-Wigner string transformation, which in 1D introduces a phase factor to maintain anticommutation along a specified direction—in this case, the direction of the vortex line. By applying this phase factor, $U_2$ enforces the anticommutative nature of fermionic operators on particles aligned along the same vortex, as would be expected for fermions confined to a line.

Interestingly, the $U_2$ phase factor corresponds to a unitary $U(1)$ gauge transformation when particle density $n_{\mathbf{r}}$ is continuous. Although it modifies the phase of fermionic operators along the $z$ axis, it does not produce any net magnetic flux across the system, effectively leaving the physics of the system unchanged in terms of gauge interactions. 
To confirm that this gauge transformation does not create a physical magnetic field, one can take the continuous limit of the trivial identity product
$\prod_i U^+_2({\bf r}_i) U_2({\bf r}_{i+1}),$
which becomes the Wilson loop $C$ of the 3D gauge vector potential $U_2^+ \partial U_2$
\bea
\label{WL}
1&=&\prod_i U^+_2({\bf r}_i) U_2({\bf r}_{i+1}) \\
&=&\text{Tr}\left(\mathcal{P} \exp \left\{i\oint_C U^+_2({\bf r}) \partial U_2({\bf r}) \right\}\right),\nn
\ena
where $\mathcal{P}$ is the path-ordering operator. The fact that the Wilson loop evaluates to unity implies that this phase transformation does not create any enclosed flux around a closed path $C$, confirming that the transformation does not alter the physical gauge field. This property is similar to the behavior observed in the XX spin chain after the Jordan-Wigner transformation, where the product of spin-1/2 raising and lowering operators, $S^{\pm}$, on neighboring sites transform into fermionic operators, $c^{\pm}$, without affecting the form of the Hamiltonian. This invariance ensures that the physical properties of the system are retained after the transformation, with the transformation itself solely altering the statistics of the operators.

The CVLL state we discussed above serves as a variational state designed to more efficiently lower the energy of low-density interacting bosons with moat dispersion, compared to Bose condensates. It also aims to capture the essential statistical and phase properties that are characteristic of fermionized bosons through CS flux attachment. In the next section, we will explore the topological field theory that underpins this construction, identifying it as the theory that yields the CVLL state from the Euler-Lagrange equations of motion.

\section{Field theory of the chiral vortex-line liquid}

The CS flux attachment procedure in two spatial dimensions enables a hard-core bosonic system to be transformed into an equivalent description of spinless fermions coupled to the $U(1)$ CS gauge field, termed a statistical CS action. This can be derived by noting that the fermionic action resulting from this transformation is coupled to a fixed CS gauge field, where the associated CS vector potential is determined by the local density of fermions, reflecting the flux attachment. To ensure the consistency of this transformation, one introduces a fluctuating gauge field, achieved by enforcing the Gauss law with a Lagrange multiplier. This gauge field effectively mediates interactions in the fermionic action, making it equivalent to the original bosonic action. The resulting fermionic theory is then coupled to a dynamic $U(1)$ gauge field characterized by the statistical CS action. Our objective here is to extend this program to 3D by implementing the 3D version of the flux attachment discussed in the previous section.  

An important novel aspect of our ``CS theory" in three spatial dimensions is the presence of a variational function, $f(z)$, which defines a continuous form of the vortex line and serves as a degree of freedom for minimizing the energy of the system. To explore the field theory corresponding to the 3D bosonic wavefunction in Eq.~(\ref{CS-3}), let us consider the quantum CVLL state of particles located on distinct vortex lines, represented by $q_{\bf l}(z)$ and $q_{\bf l^\prime}(z)$ with ${\bf l} \neq {\bf l^\prime}$, assuming that no pair of particles reside on the same vortex line. This assumption is particularly useful because our goal is to derive the CS gauge field from this ansatz while keeping in mind that JW strings along vortex lines do not produce any gauge fields. To this end, we consider the following bosonic wave function

\begin{align}
\label{CS-4} 
&\hat\Psi_B({\bf r}_i)= \\
&\exp\Big\{i m \sum_{j\neq i}\arg\big[q^{}_i(z_i)-q^{}_j(z_j)\big]\hat n_j \Big\}\hat\Psi_F({\bf r}_i). \nn
\end{align}

The phase factor in the transformation from bosonic to fermionic wavefunction introduces a gauge-fixed CS-type gauge field. This modification allows us to express the Hamiltonian of the model in terms of fermionic matter fields that are coupled to the $U(1)$ CS-type spin-1  field, resulting in inducing a covariant derivative, i.e., $i \partial_a \rightarrow i \partial_a +A_a({\bf r})$
with $a=x,y,z$. Such a transformation creates a $U(1)$ vector potential
\bea
\label{A-1}
A_a({\bf r})&=&m\partial_a \sum_{{\bf l'}\neq {\bf l},z^\prime} \log\Bigg[\frac{q^{}_{\bf l}(z)-q^{}_{\bf l'}(z')}{\mid q^{}_{\bf l}(z)-q^{}_{\bf l'}(z')\mid}\Bigg],\nn\\
\ena
where $a=x,y,z$, around the vortex line located at ${\bf r}=({\bf l},z)$ with ${\bf l}=(x,y)$. This field is singular on a vortex line, $q^{}_{\bf l}(z)$, and is attached to a fermionic field. A direct calculation of the curl,
 $H^a({\bf r})=\epsilon^{abc} \partial_b A_c({\bf r})$, created by the vortex line, gives the effective CS magnetic field with components
 \bea 
 \label{B-1}
 H^x({\bf r})&=&-2\pi m x \frac{f^{\prime}(z)}{f(z)} n_{{\bf r}}=2\pi h^x  n_{{\bf r}}
 \nn\\
  H^y({\bf r})&=&-2\pi m y \frac{f^{\prime}(z)}{f(z)} n_{{\bf r}}=2\pi h^y  n_{{\bf r}}
  \\
   H^z({\bf r})&=&2\pi m n_{{\bf r}}=2\pi h^z  n_{{\bf r}},
   \nn
 \ena
where $n_{{\bf r}}= L \sum_{z^\prime} \sum_{{\bf l}^{\prime} \neq {\bf l}}\delta(q^{}_{\bf l}(z)-q^{}_{{\bf l}^{\prime}}(z^\prime))$ is the density of vortex lines at ${\bf r}$ while $L$ is their total number. Here, for the sake of simplification, we introduce a new notation, $h^{a}$, defined as $h^x=-mx f'(z)/f(z)$, $h^y=-my f'(z)/f(z)$, and $h^z=m$.


In 3D space, the emergence of a statistical gauge field is more subtle than in 2D systems. In 2D, the statistical CS term in action introduces a CS gauge field which, in the constant-density approximation, leads to an emergent magnetic field given by the curl of a vector potential. This field effectively captures the statistical interactions in the system. One encounters fundamental differences in our fermionization approach to 3D interacting boson systems. The statistical term in the action resulting from the 3D fermionization procedure leads to an emergent magnetic field. However, when $ f(z) \neq 1$, this field does not satisfy the condition of being the curl of a conventional vector potential in a flat space under the constant particle density approximation. This contrasts with the 2D case, where the emergent field remains expressible in terms of a CS gauge potential. Mathematically, in 3D, the emergent magnetic field Eq. (\ref{B-1}) in a flat space generally does not satisfy ${\bf H}={\bf \nabla} \times {\bf A}$ in the constant density limit. The failure of this condition arises due to the intrinsic nature of 3D gauge theories, where the induced field does not map straightforwardly to an Abelian-like flux attachment as it does in 2D. Instead, the emergent vortex excitations do not admit a simple curl representation. However, if one allows for a nontrivial spatial curvature, characterized by a position-dependent metric function defined in terms of $f(z) \neq 1$, the emergent magnetic field in the constant density approximation can indeed be expressed as the curl of a modified vector potential in this curved space. This suggests that the statistical gauge structure in 3D is intimately connected to the geometry of the underlying space, unlike in 2D, where it remains well defined in a flat space. In such curved backgrounds, the emergent vector potential $\mathbf{A} $ is modified by metric-dependent terms, and the magnetic field acquires additional contributions coming from the curvature. This allows the emergent field to be rewritten in the form: $ \mathbf{H} = \nabla_g \times \mathbf{A} $ where $ \nabla_g $ is the covariant derivative associated with the curved spatial metric. The function $ f(z) $ effectively encodes deviations from a flat space, and when properly incorporated, it restores the expressibility of the emergent field as the curl of a vector potential. 

The dependence on curvature suggests that the emergent gauge fields in 3D statistical gauge theories may be linked to gravitational effects. 


In the fermionization approach to spin systems~\cite{sedrakyan-2012,sedrakyan-2015-1, sedrakyan-2015-2}, one typically resorts to a mean-field approximation based on the homogeneous density of particles or vortices, leading to a constant CS magnetic field.
From Eq. (\ref{B-1}), it is clear that with the vector potential ${\bf H}={\bf \nabla} \times {\bf A}$, we cannot have a constant magnetic field based on the constant density of fluxes.
However, it can be fixed by introducing a manifold $\mathcal{M}$ with a non-dynamical metric $g_{ab}^{(3)}$ which is independent of time, under which the components of vectors $\mathbf{A}$ and $\mathbf{H}$ in the orthonormal basis become $\mathcal{A}^{\bar{a}}$ and $\mathcal{H}^{\bar{a}}$ with $\bar{a}=1,2,3$.
The general relation between ${\cal A}^{\bar{a}}$ and ${\cal H}^{\bar{a}}$ is
\begin{equation} \label{B-2}
\begin{aligned}
   \mathcal{H}^{d} = \mathcal{H}^{\bar{b}} \mathbf{e}_{\bar{b}}^{~d} = \frac{1}{\sqrt{g^{(3)}}} \epsilon^{dab} \nabla_{a} (g^{(3)}_{bc} \mathcal{A}^{\bar{a}} \mathbf{e}_{\bar{a}}^{~c}).
\end{aligned}
\end{equation}
where
$\mathbf{e}_{\bar{a}}^{~a}$ is the deribein (vielbein in 3D) satisfying
$g^{(3)}_{ab} \mathbf{e}^{~a}_{\bar{a}} \mathbf{e}^{~b}_{\bar{b}} = \delta_{\bar{a}\bar{b}}$,
the affine connection acts on a vector $V^{a}$ as $\nabla_{a} V^{b} = \partial_{a} V^{b} + \Gamma^{b}_{~ac}V^{c}$ with $\Gamma^{b}_{~ac}$ being the Christoffel symbols, and  $g^{(3)}$ is the determinant of the metric $g_{ab}^{(3)}$.Throughout the paper, we used the convention that Latin letters $a,b,c,...$ are 3D Euclidean space indices,
barred letters $\bar{a},\bar{b},\bar{c},...$are the coordinate chart indices of the 3D curved space,
and Greek letters $\mu, \nu, \rho, ...$ are the coordinate chart indices of the (3+1)D curved space.

The problem is now finding a suitable metric $g^{(3)}_{ab}$ for the new set of vector potentials ${\cal A}(\Vec{x}')$ residing on ${\cal M}$.
The simplest choice is to set ${\cal A}^3=0$ and assume that the metric $g^{(3)}_{ab}$ is diagonal with the form $\text{diag} \{ Q,Q,1 \}$.
With this choice,
one can solve Eq. (\ref{B-2}) using the magnetic field described in (Eq. (\ref{B-1})), resulting in the following expression for ${\cal A}(\Vec{x})$
 \bea
 \label{A-2}
 &{\cal A}^1=-y\pi m n_r(\log f(z)+1),\\ \nn
 &{\cal A}^2= x \pi m n_r(\log f(z)+1),\\ 
 &{\cal A}^3=0.\nn
 \ena
In this equation $|g|=(\log f(z)+1)^4$ is the determinant of the new curved metric $\hat{g}$
\bea
\label{metric}
\hat{g}=\left( 
\begin{array}{cccc}
	-1&0&0&0\\
	0&(\log f(z)+1)^2&0&0\\
	0&0&(\log f(z)+1)^2&0\\
	0&0&0& 1
\end{array}
\right).
\ena

\vspace{0.3cm}

 Can we formulate a gauge field theory as a solution of the Euler-Lagrange equations of motion that will reproduce expressions Eq. (\ref{A-1}) and Eq. (\ref{B-1}) as Eq. (\ref{B-2}) and Eq. (\ref{A-2}), respectively? This can be done by considering a fermionic system in the Galilean gravitational field. In general, the concrete form of the kinetic energy of fermions can be any relativistic Dirac-type with non-relativistic quadratic terms. The main property of the system we will look into is that its single-particle dispersion must be degenerate on a 2D closed manifold embedded in a 3D reciprocal space as defined by the kinetic energies in Eq. (\ref{kinetic}). For simplicity, consider
\bea
\label{Hkin}
K({\bf k})=\int d^3\Vec{x} \sum_\mu \Psi^{B \dagger}_{\tau{\bf r}} \: \hat{K} \: \Psi^{B}_{\tau{\bf r}},
\ena
where $\Psi^{B}_{\tau{\bf r}}$ is a bosonic wave function in (3+1)D with imaginary time component $\tau$. 
Using CS transformation, Eq.~(\ref{CS-4}), as a gauge transformation for transforming bosonic wavefunction $\Psi^{B}_{\tau{\bf r}}$ into fermionic wavefunction $\Psi_{\tau{\bf r}}$ with metric $ g_{\mu\nu}$
which has a conformal (Weyl) form in the $(x,y)$ plane coming from the covariant part of the Lagrangian, we come to the action 

\bea
\label{S-1}
S&=&\int d\tau d^3\Vec{x} \sqrt{g}\\
&\times&\Bigg\{\Psi^+_{\tau{\bf r}}\Bigg((\partial_{\tau}-{\cal A}_0)+K(|\hat{k}|)\Bigg)\Psi_{\tau{\bf r}}\nn\\
&+&\frac{i m}{2\pi}
{\cal B}^{\bar{b}}
\frac{1}{\sqrt{g^{(3)}}} \epsilon^{dab} \nabla_{a} (g^{(3)}_{bc} \mathcal{A}^{\bar{a}} \mathbf{e}_{\bar{a}}^{~c}) \mathbf{e}^{d}_{~\bar{b}}
\Bigg\}. \nn
\ena
where $a, b,c,d = x,y,z$ and $\bar{a}, \bar{b} = 1,2,3$ are the spatial coordinates, and $\tau$ is the imaginary time defined by a Wick rotation as $\tau=-it$. In Eq. (\ref{S-1}) $K$ is the kinetic energy written as in Eq. (\ref{kinetic}), and the momentum operator in the curved space is defined as $|\hat{k}|=\frac{1}{\sqrt{\pi}}\hat{k}^2\int_{-\infty}^{\infty}ds e^{-s^2\,\hat{k}^2}$ where $\hat{k}^2=\frac{1}{\sqrt{-g}}D_{\bar{a}}(\sqrt{-g}g^{\bar{a}\bar{b}}D_{\bar{b}})$ with the covariant derivative defined as $D_{\bar{a}}=\nabla_{\bar{a}}-i \mathcal{A}_{\bar{a}}$, and $\nabla_{\bar{a}}$ being the affine connection. Here, we define a fluctuating 4--vector potential as ${\cal A}_{\mu}=({\cal A}_0,{\cal A}_{\bar{a}})$, where ${\cal A}_0 =\sum_{a=x,y,z}{\cal B}_a h^a$ with a Lagrange multiplier ${\cal B}^{\mu}$ acquired by imposing Eq. (\ref{B-1}). ${\cal A}_0$ can be interpreted as a temporal component of the vector potential ${\bf {\cal A}}$, therefore, we created a (3+1)D action on a curved space-time.
In this action, the fermions are spinless, so we do not need a spin connection.

In our approach, we derive the statistical term in the action (\ref{S-1}) by implementing a three-dimensional fermionization procedure, which combines a two-dimensional Chern-Simons transformation in the plane with a Jordan-Wigner transformation along vortex lines that extend in the third dimension. This construction is carried out in a fixed gauge: we choose a centrally symmetric gauge for the Chern-Simons field in the plane and define the Jordan-Wigner strings along specified vortex lines. The resulting statistical term in the effective action is consequently not gauge invariant. Nevertheless, similar to the conventional two-dimensional Chern-Simons fermionization framework, one can relax this fixed-gauge constraint by formulating the transformation in an arbitrary gauge, thereby promoting the theory to a gauge-invariant U(1) quantum field theory. Such a reformulation would restore gauge invariance, which is often crucial to ensuring the renormalizability and consistency of the theory. However, for the present work, we proceed within the fixed-gauge framework without invoking gauge invariance, leaving this as an open question for the future, along with renormalizability considerations.

\def\mean#1{\left< #1 \right>}

To analyze the low-energy fluctuations of the spin-1 vector field ${\cal A}$, we will first integrate out the fermionic field. This step allows us to construct an effective action that captures the interactions and dynamics of the field, isolating it from the influence of the fermions. Following this, we will perform a series expansion of the effective action around the mean-field $\langle{\delta \cal A} \rangle$ value, retaining terms up to the second order in the fluctuation of the gauge field ${\delta \cal A}$. This expansion is expected to capture the primary contributions to the low-energy behavior of the gauge field while simplifying the analysis.
Unlike a traditional approach that expands around a saddle-point configuration, where the field configuration minimizes the action, our method does not assume that the field configuration is near a minimum. Expanding around a saddle point is more restrictive, as it is typically effective only when the fermionic field's behavior is approximately quadratic in fluctuations around the minimum. However, by expanding around the mean-field value, we can avoid any limitations related to the shape of the potential at the saddle point. This is particularly beneficial in cases where the fermionic field does not exhibit a quadratic structure.
Moreover, this around-mean-field-point expansion naturally accommodates potential Goldstone modes, if such exist, which would otherwise complicate or limit the validity of an expansion around a saddle point. The details will be presented in the next section for the representative case of the cylindrical moat band.

\section{Low-energy effective field theory for interacting bosons with cylindrical moat-band dispersion}

 \begin{figure}[t]
    \centering
    {\includegraphics[width=0.45\textwidth]{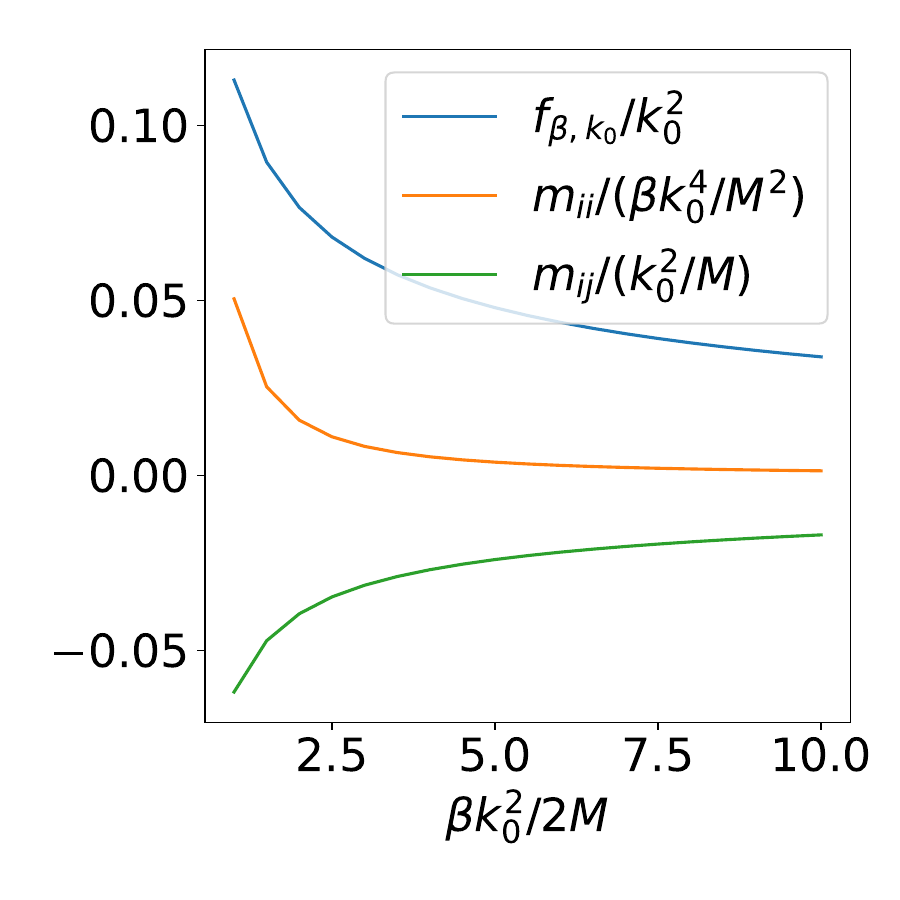}}
    \caption{\label{mass}
    Dimensionless parameters in the polarization operators as a function of the dimensionless inverse temperature $\frac{\beta k_0^2}{2M}$, where $m_{ii}$ is the $m_{xx}$ or $m_{yy}$ and $m_{ij}$ is the $m_{xy}$ and $m_{yx}$.
    }
\end{figure}

In this section, we develop a low-energy effective field theory for interacting bosons with a cylindrical moat-band dispersion, focusing on the properties of low-energy excitations emerging from CVLL. The path integral formulation leads us to an effective action where integrating out fermionic degrees of freedom reveals the nature of fluctuating fields around the mean-field solution. Specifically, we derive the polarization operator within the Random Phase Approximation (RPA), capturing thermal and quantum fluctuations that define the nature of the low-energy excitations. 

Due to the translational symmetry in the $z$ direction, we take $f(z) = 1$. With this simplification, the path integral partition function of Eq. (\ref{S-1}) becomes
\bea
\label{partition}
{\cal Z}=\int  {\cal D}\Psi^{\dagger}\,{\cal D}\Psi\,{\cal D}{\cal A}_{\mu} \;e^{-L_z S_c[\Psi^{\dagger}, \Psi,{\cal A}_{\mu}]}
\ena
with
\begin{equation}
\begin{aligned}
    S_c = \int d\tau dxdy \Bigg\{& \Psi^+_{\tau{\bf r}}\Bigg((\partial_{\tau}-{\cal A}_0)+K_c(|\hat{k}|)\Bigg)\Psi_{\tau{\bf r}}\\
    &+ \frac{i m}{4\pi} \sum_{\mu,\nu,\rho=\tau,x,y} \epsilon_{\mu\nu\rho} \mathcal{A}_\mu \partial_\nu A_\rho \Bigg\}
\end{aligned}
\end{equation}
Since the action is independent of $z$, we have integrated the $z$ direction $\int dz \to L_z$. Moreover, we imposed $\nabla \cdot \mathcal{A} = 0$ in 2D, so that the term $\mathcal{A}_0 \sum_{a,b=x,y} \epsilon_{ab} \partial_a \mathcal{A}_b$ becomes the CS term up to a total derivative.
In order to get an effective action, we need to integrate out the fermions first
\bea
\label{partition_eff}
{\cal Z}=\int {\cal D}{\cal A}_{\mu} \;e^{-S_{\text{eff}}[{\cal A}_{\mu}] + L_z S_{\rm CS}[{\cal A}_{\mu}]}
\ena
with
\begin{align}
\label{effective}
S_{\text{eff}}[{\cal A}_{\mu}]= - \log \det (\partial_{\tau}-{\cal A}_{0}+K(|\hat{k}|))
\end{align}
where a constant term $- \tr \log L_z$ has been dropped in $S_{\text{eff}}$.

As discussed in the appendix \ref{polarizationdef}, $S_{\text{eff}}[\mathcal{A}]$ contributes two terms: the mean field contribution giving rise to an extra CS term and the fluctuation around the mean-field $S_{0,\text{eff}}[\delta\mathcal{A}]$ which can be computed from the one-loop and random phase approximation (RPA). To calculate the one-loop polarization operator for spinless fermions, we just need to expand the determinant with respect to the fluctuating statistical gauge field $\delta {\cal A}$.

Within the RPA, one can compute the polarization operator at finite temperature from the effective action Eq.~(\ref{effective}) at low energy ($|{\bf q}| \ll k_0$) as detailed in the appendix \ref{polarizationdef}. The resultant polarization operator consists of two parts, the first part comes from the effective action, and the second term comes from the CS action as follows
\begin{widetext}
\begin{equation}\label{Pi}
\begin{aligned}
    \Pi_{\mu\nu}(\mathbf{q}) =
    \frac{f_{\beta,k_0}^2 M({\bf q})}{(q_x^2+q_y^2)^2 \left( \beta f_{\beta,k_0} M({\bf q}) + (m_{xx}^2-m_{xy}^2) q_{\tau}^2 \right)}
    \begin{pmatrix}
        q_{\tau }^2 & q_{\tau } q_x & q_{\tau } q_y \\
        q_{\tau } q_x & q_x^2 & q_x q_y \\
        q_{\tau } q_y & q_x q_y & q_y^2
    \end{pmatrix}
    +
    \frac{(m+1) L_z}{2\pi}
    \begin{pmatrix}
        0 & 0 & 0 \\
        0 & 0 & \frac{q_{\tau}^2+q_x^2+q_y^2}{q_{\tau}} \\
        0 & -\frac{q_{\tau}^2+q_x^2+q_y^2}{q_{\tau}} & 0
    \end{pmatrix},
\end{aligned}
\end{equation}
\end{widetext}
\mbox{}\par\clearpage
where
\begin{equation}
\begin{aligned}
    f_{\beta, k_0} &= k_0^2 \int_{0}^{\infty} \frac{dx}{2\pi} \frac{x}{2\cosh(\frac{\beta k_0^2}{2M} (x - 1)^2) + 2},\\
    m_{xx} &= \frac{\beta k_0^4}{M^2} \int_{0}^{\infty} \frac{dx}{2\pi} \frac{\frac{1}{2} x(x-1)^2}{2\cosh(\frac{\beta k_0^2}{2M} (x - 1)^2) + 2}, \\
    m_{xy} &= \frac{k_0^2}{M} \int_{0}^{\infty} \frac{dx}{2\pi} \frac{\frac{1}{4} - \frac{3}{4}x}{2\cosh(\frac{\beta k_0^2}{2M} (x - 1)^2) + 2},
\end{aligned}
\end{equation}
are parameters depending on the inverse temperature $\beta$ and moat radius $k_0$, $M({\bf q}) = m_{xx} (q_x^2+q_y^2) - 2 m_{xy} q_x q_y$, and $L_z$ which is the system size in the $z$ direction. The temperature dependence of these parameters is shown in Fig. \ref{mass}.

At zero temperature, the parameters $ f_{\beta, k_0} $, $ m_{xx} $, and $ m_{xy} $ vanish, causing the first term in the polarization operator to disappear. At low temperatures, when $ \beta^{-1} \lesssim k_0^2 / 2M$, these parameters behave as the square root of the temperature:  
\begin{eqnarray} \label{asymptotes}
&&\frac{f_{\beta, k_0}}{k_0^2} \simeq \frac{1.34744}{4\pi}  \sqrt{\frac{2M}{\beta k_0^2}}, \nn\\
&&m_{xx} \simeq \frac{1.07215}{2\pi} \sqrt{\frac{k_0^2}{2M\beta}}, \nn\\
&&m_{xy} \simeq -\frac{1.34744}{4\pi} \sqrt{\frac{k_0^2}{2M\beta}},
\end{eqnarray} 
where the small and subleading terms in powers of $\frac{M}{\beta k_0^2}\lesssim 1$ are dropped. These asymptotes indicate that the temperature-dependent term in the polarization operator at low temperatures vanishes singularly, as $\sim\beta^{-3/2}$. This fact suggests a continuous phase transition at zero temperature, $\beta^{-1}\rightarrow 0$.

Importantly, independent of temperatures, the characteristic equation for the polarization operator (Eq. (\ref{Pi})) leads to a dispersion for the $ x $- and $ y $-components of the fluctuating vector field, forming a light cone: 
\begin{equation} 
\label{reldis}
\omega = \pm\sqrt{q_x^2 + q_y^2}.
\end{equation}

\begin{figure}[t]
\centering
\includegraphics[width=0.3\textwidth]{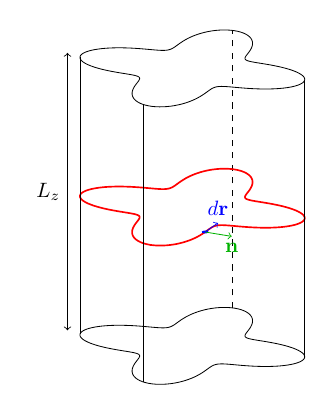}
\caption{Schematic diagram for a 3D cylindrical sample of arbitrary shape in real space. The red curve is the boundary of a 2D cut at a fixed value of $z$.}
\label{fig:boundary}
 \end{figure}
 
 To support a Dirac cone dispersion Eq.~(\ref{reldis}), the complete low-energy effective theory must exhibit an emergent relativistic invariance, specifically Lorentz invariance in (2+1)D. This is indeed the case, and importantly, this relativistic invariance is an emergent property of the low-energy theory of low-energy excitations. We note that the original theory in Eq.~(\ref{S-1}) lacks relativistic invariance due to the nonrelativistic nature of the moat-band dispersion. The reason is that the original bosonic matter fields with the moat dispersion are gapped, lying far beyond the low-energy description, indicating the emergence of collective modes at low energies and low temperatures.

To further understand the low-energy effective field theory of the CVLL state, we focus on the effective action at zero temperature in real space
\begin{eqnarray}
\label{effective1}
    &&S_{0,\rm eff} = \frac{(m+1) L_z}{2\pi}\times \\
    &&\sum_{i,j=x,y}
    \int dt\int dxdy \epsilon_{ij} \delta\mathcal{A}_i \left( \partial_t + \frac12 \text{sgn}(t) \nabla^2 \right) \delta\mathcal{A}_j,\nonumber
\end{eqnarray}
with $\delta\mathcal{A}$ being the spin-1 vector field fluctuation around the mean field value.
To see its topological nature, one can perform the integral by parts for the second term, which gives a boundary term and an anti-symmetric bulk term that vanishes under summation as detailed in \cref{boundary}
\begin{equation}
\begin{aligned}
\label{effbound}
    S_{0,\rm eff} = \frac{(m+1) L_z}{2\pi} \sum_{i,j=x,y} \int dt \Big(& \int dxdy \epsilon_{ij} \delta\mathcal{A}_i \partial_t \delta\mathcal{A}_j \\
    -\frac{1}{2} \text{sgn}(t) & \oint dr\;\mathbf{n} \cdot \epsilon_{ij} \delta\mathcal{A}_i \nabla \delta\mathcal{A}_j \Big),
\end{aligned}
\end{equation}
where the second term is the integral over the boundary of the system with $\bf n$ being the normal vector pointing outside the system as labeled in Fig. \ref{fig:boundary}.

The free low-energy effective action Eq.~(\ref{effbound}) consists of two terms, $S_{0,\rm eff}=S_{0, \rm bulk}+S_{\rm bound}$, a free bulk term $S_{0, \rm bulk}=\frac{(m+1) L_z}{2\pi} \sum_{i,j=x,y} \int dt \int dxdy \; (\delta\mathcal{A}\times \partial_t \delta\mathcal{A})_z$ and the boundary term that can be rewritten in the following form
\begin{equation}
    S_{\text{bound}}=-\frac{(m+1) L_z}{2\pi} \int dt\; \text{sgn}(t)\oint dr\;({\bf n}\times{\bf \delta\mathcal{A}})_z({\bf \nabla}\delta\mathcal{A}).
\end{equation}
The free action $S_{0,\rm eff}$ is fully invariant under continuous SO(2) rotations in the $(x,y)$ plane, but it breaks time-reversal (T) and reflection (P) symmetries. Combined PT symmetry is preserved in this action, indicating a characteristic feature common to topologically ordered systems supporting chiral edge states. In the following, we argue that the latter is indeed the property of the CVLL phase.

The bulk action $S_{0, \rm bulk}$ has the form of a topological theory with a first-order derivative in time, analogous to a two-dimensional CS-type action, but notably without spatial derivative terms. 
The action $S_{0, \rm bulk}$ lacks explicit energy dispersion terms (no spatial gradients), thus, it supports a flat band for the fluctuating vector field components. The system has infinitely degenerate static solutions parameterized by time-independent fields $\delta\mathcal{A}_i(x,y)$, with $\delta\mathcal{A}_x(x,y)$ and $\delta\mathcal{A}_y(x,y)$ being the conjugate variables. In other words, we deal with the flat-band situation for quasiparticle excitations in the bulk, which are localized in real space and exhibit an absence of propagation. The theory describes a purely topological, zero-energy sector where bulk modes do not propagate.

Going beyond the quadratic terms in the expansion in Eq. (\ref{effective}), a quartic interaction term explicitly introduces energy scales and spatial variations. The original flat-band, containing only a time derivative term, would now gain spatial dependence through potential gradient terms upon quantum fluctuations.
The presence of such nonlinear terms could lead to the emergence of low-energy excitations and spatial correlations. Low-energy excitations from these flat-band localized vacuum states could emerge as collective modes. Since the interaction term, evaluated from further expanding Eq. (\ref{effective}), is positive definite, we conjecture that the quasiparticle excitations will be gapped, thereby preserving the $SO(2)$ rotational symmetry of the Hamiltonian. This also suggests the stability of our mean-field solution and the emergence of vortex-line excitations upon restoring the $z$-dimension in the problem. The braiding statistics of the emergent vortex-line excitations is an open problem we propose to study by investigating the interaction terms and their effect on the flat-band localized states in the bulk. 

The boundary action $ S_{\text{bound}}$ represents a quadratic Lagrangian for a two-component vector field $\delta\mathcal{A}_i(x,y)$, with $i=x,y$, defined on a (1+1)D boundary of the 2D slice of the system at a fixed value of $z$ (see Fig. \ref{fig:boundary}). The boundary action breaks the individual T and P symmetries while preserving the combined PT symmetry. Upon reintroducing the time derivative, $\partial_t$, the relativistic dispersion $\omega\sim |{\bf q}|$ emerges, characterizing gapless chiral bosonic edge excitations, analogous to quantum Hall edge states and edge states in CSLs.

With no dependence on $\partial_z$, the frequency $\omega$ of excitations is independent of the momentum $q\strut_z$. This leads to a flat band along the $q\strut_z$ direction, indicating that there is no gap opening or velocity in the $z$-direction. Physically, this can be viewed as a stack of 2D chiral boundary modes, each localized in a plane for each value of $z$. Thus, in 3D, the same 2D chiral physics is replicated along the $z$-direction. As a result, the $1/K$ fractional vortex excitations inherent to 2D CS theories at level $K=m+1$, due to this stacking, become fractionalized vortex-line excitations (flux flips along the JW vortex lines) of the CVLL state. 

The resulting edge states are a continuum of boundary modes, {\it surface modes}, that do not disperse in $z$ and continue to break time-reversal and reflection symmetries in the same manner, slice by slice. The only new symmetry is the trivial translation invariance along $z$ (and possibly large gauge transformations if one includes them), but the action remains non-invariant under full 3D rotation.

\section{
CVLL ground state of interacting bosons with cylindrical moat dispersion: Monte Carlo simulation of the equation of state
}

\begin{figure}[t]
    \centering
    {\includegraphics[width=0.45\textwidth]{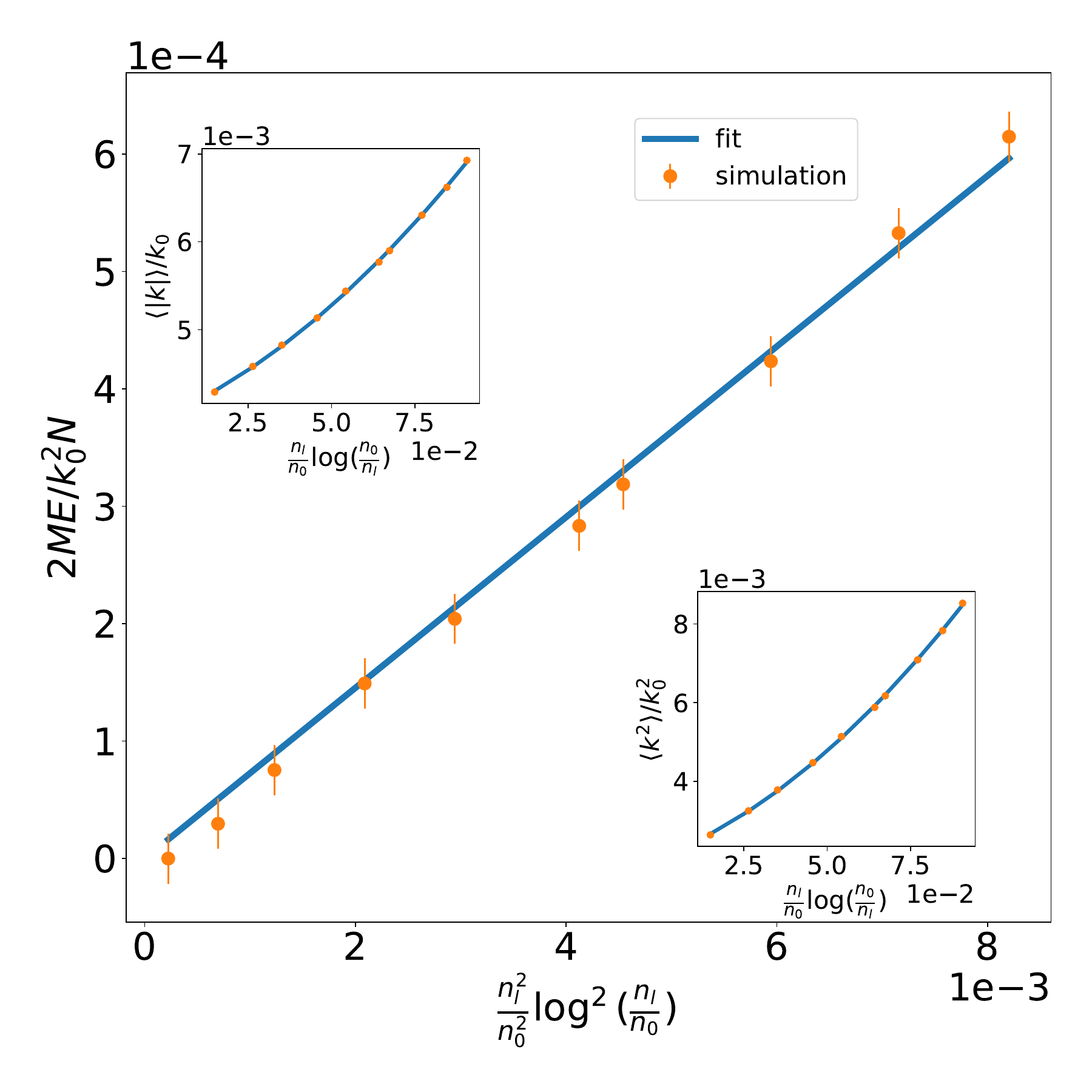}}
    \caption{\label{E_MC}
    The energy of the factorized CVLL wavefunction Eq.~(\ref{PsiB_MC}) with $m=1$. The inset shows the scaling of 
    $\langle k \rangle = \bra{\Psi_B} |k_{\text 2D}| \ket{\Psi_B}$
    and
    $\langle k^2 \rangle = \bra{\Psi_B} k_{\text 2D}^2 \ket{\Psi_B}$.
    }
\end{figure}

After discussing the stability of the mean-field theory, in this section, we study the CVLL state, focusing on the cylindrical moat band. This setup maintains translational symmetry along the $z$ spatial dimension, thereby enabling a simplified treatment of the wavefunction in terms of 2D and $z$-separable components. We will use Monte Carlo (MC) simulations to demonstrate that the CVLL state remains energetically favorable compared to condensation states, even with finite interaction strength $g_0$, when the particle density $n$ is sufficiently low. 

Because of the translation symmetry in the $z$ direction, the wavefunction can be separable, i.e., $f(z) = 1$, leading to $\hat{g} = {\rm diag}\{-1,1,1,1\}$, and it simplifies to the following form
\begin{equation} \label{PsiB_MC}
\begin{aligned}
    \Psi_B({\bf r}_1,{\bf r}_2,\cdots,{\bf r}_L) =
    \Psi_{B,2d}(w_1,w_2,\cdots,w_L) \\
    \times \Psi_{B,z}(z_1,z_2,\cdots,z_L)
\end{aligned}
\end{equation}
where $w_i = x_i + i y_i$ denotes the complex coordinates in 2D plane and
\begin{equation}
\Psi_{B,2d} =
\prod_{i < j}\frac{w_{i}-w_{j}}{| w_{i}-w_{j} |} \Psi_F(w_1,w_2,\cdots, w_L).
\end{equation}
In the above 2D wavefunction $\Psi_{B,2d}$, we choose $m = 1$ as it corresponds to the simplest case of the CVLL. Generally, the fermionic part $\Psi_F(w_1,w_2,\cdots, w_L)$ can be any anti-symmetric function. However, we will focus on the case where the smeared density is spatially uniform, which is valid in the low-density limit. Under this assumption, the gauge flux becomes uniform, and the effective action (\ref{S-1}) simplifies to an action that describes free fermions in a uniform gauge flux. Hence, $\Psi_F(w_1,w_2,\cdots, w_L)$ is a Slater determinant of Landau levels
\[\Psi_{F}^{(l)}\left(z_{1},z_{2},\ldots,z_{N}\right)=\frac{1}{\sqrt{N!}} \det_{\substack{-l\leq m\leq N-l-1\\1\leq j\leq N}} \left[\chi_{m}^{(l)}\left(z_{j}\right)\right],\]
where each $\chi_{m}^{(l)} (z)$ represents the single-particle Landau level wavefunction in a uniform gauge field fixed in the symmetric gauge
\begin{align}
 &\chi_{m}^{(l)} (z)=\\
&\begin{cases}
        \frac{(-1)^{l}\sqrt{l!}}{l_{B}\sqrt{2\pi2^{m}(l+m)!}}\left(\frac{z}{l_{B}}\right)^{m}e^{-\frac{|z|^{2}}{4l_{B}^{2}}} L_{l}^{(m)}\left[\frac{|z|^{2}}{2l_{B}^{2}}\right], \hspace{15pt} m\geq0\\
        \frac{(-1)^{(l+m)}\sqrt{(l+m)!}}{l_{B}\sqrt{2\pi2^{-m}l!}}\left(\frac{\bar{z}}{l_{B}}\right)^{-m}e^{-\frac{|z|^{2}}{4l_{B}^{2}}} L_{l+m}^{(-m)}\left[\frac{|z|^{2}}{2l_{B}^{2}}\right], m\leq0
    \end{cases} \nn
\end{align}
where $L_l^{(m)}(x)$ is the generalized Laguerre polynomial with $l_B = 1/\sqrt{2 \pi n_l}$ being the magnetic length.

To determine the ground state properties of the CVLL phase, we first consider the mean-field energy $E$ in the presence of a uniform gauge field. For particles in this configuration, the energy is given by:
\begin{equation}
E=\frac{1}{2M}\Big(\sqrt{(2l+1) M\omega_{c}}-k_{0}\Big)^{2}
\end{equation}
where $\omega_c = 2 \pi n / M$ is the cyclotron frequency in terms of the particle density $n$~\cite{sedrakyan-2015-2}. To get discrete values of $n$ for different $l$, we must minimize the energy 
\begin{equation} \label{nl}
    n_l=\frac{k_0^2}{2\pi(2l+1)},\quad l\in\mathbb{N}.
\end{equation}
with a large Landau level index $l$, the optimized $n_l$ becomes small and almost continuous.

At low density, the energy $E_{3D} = \bra{\Psi_B} K_S \ket{\Psi_B} = \bra{\Psi_B} k_{\text 2D}^2 -2 k_0 |k_{\text 2D}| + k_0^2 \ket{\Psi_B}$ of the CVLL state is reduced to 2D and the energy per particle scales as $n^2 \ln^2(n)$ shown by the previously established 2D CSL studies~\cite{wei2023chiral,sedrakyan-2015-2}.
To clarify this in the CVLL state, a Monte Carlo simulation of the energy integral is performed. For this simulation, we set the particle number $N = 60$ and vary the Landau level index $l = 20, 22, 25, 30, 32, 40, 50, 70, 100, 200$, which also determines the density according to Eq.~(\ref{nl}). The results are summarized in Fig.~\ref{E_MC}, where we observe the CVLL state's energy per particle scaling as $n^2 \ln^2(n)$ at low density, as expected.
This energy should be compared to the energy of the condensation $\frac{E_c}{N} \sim g_0 n$. The MC results demonstrate that the CVLL state's energy remains lower than that of the condensate $E_c$, particularly in the low-density regime where the interaction strength $g_0$ is finite. This finding confirms that the CVLL state is energetically favored over BEC for low densities, highlighting its robustness as a ground state in the cylindrical moat band systems.

To summarize this section, we construct a first-quantized CVLL wavefunction for a 3D system of $N$ bosonic particles, formulated as a 2D CS transformed state embedded within a three-dimensional space, where vertical Jordan-Wigner strings encode the third spatial dimension. In the strictly 2D limit, where the $z$ direction is absent, this state corresponds to the well-known CSL phase. This phase can be understood as a fermionic Slater determinant wavefunction that describes a fully-filled Landau level, incorporating CS phase factors that add to, rather than cancel, the phase structure of the lowest Landau level wavefunction. In this framework, the emergent CS gauge field effectively plays the role of the external magnetic field, which is proportional to the local particle density. Extending this construction to (3+1)D, we observe that due to the fermionic nature of the transformed CVLL state, its interaction energy vanishes exactly. Consequently, the total energy of the system is determined solely by its kinetic energy. Upon employing the Monte Carlo method to evaluate the kinetic energy, we obtained an explicit estimate of the total energy of the CVLL state. We showed that in the case of the cylindrical moat, the energy of the CVLL state rules out the possibility of forming a BEC at low densities, implying that the CVLL is a ground state of the system. 

The 2D CSL ground state of interacting bosons in a moat-band dispersion and the 3D CVLL ground state of bosons with a cylindrical moat surface share an identical scaling form of their equation of state, $\mu(n) \sim n^2\bigl(\ln n\bigr)^2$, where $\mu$ is the chemical potential and $n$ the boson density.  
 However, despite the similar equation of state signature, the underlying quantum orders in 2D and 3D are fundamentally distinct. In 2D, the CSL is characterized by a fully gapped topological order supporting semionic anyons, with the many-body wavefunction constructed via a Chern–Simons flux attachment that binds vortices to fermions in a single plane.  Its ground-state degeneracy and modular $K$ matrix reflect the nontrivial braiding statistics of pointlike excitations, and the entanglement spectrum exhibits a chiral conformal edge mode.
In contrast, the 3D CVLL state of bosons with cylindrical moat dispersion in momentum space leads to a condensation of vortex loops rather than point vortices.  The 3D wavefunction can be formally viewed as a stack of 2D CSL layers whose semion excitations are bound into continuous vortex lines threading the bulk.  These vortex lines form a liquid with line-like topological defects, and the system supports loop–loop and loop–point braiding statistics distinct from purely 2D anyons.  Thus, the nature of the long-range entanglement in 3D differs from that of the 2D CSL.

Thus, while the chemical-potential scaling is identical in both cases, the 2D CSL and 3D CVLL represent fundamentally different quantum phases: the former is a 2D topological order of semions, and the latter is a 3D loop-condensed state with vortex-line excitations and higher-dimensional braiding structure.

\section{Outlook} 

We have generalized the CS transformation to three spatial dimensions and formulated the fermionization/anyonization of hard-core bosons. The transformation and subsequent flux-smearing approximation define a novel state of quantum matter with emergent vortices inducing a metric in the low-energy sector of the theory. Our studies indicate that the vortices have nontrivial braiding and loop statistics and can be regarded as anyonic vortices.  Furthermore, we showed that at low densities of particles, the three-dimensional analog of the two-dimensional CSL, namely the CVLL state, is preferable against the condensate phase for short-range interacting bosons.

The results of the present work have implications for the physics of ultracold-atom systems, frustrated quantum magnets, the physics of excitons in 3D, $^4$He, and heavy-ion collisions, which are discussed below.

In the context of ultracold atoms, Ref.~\cite{Galitski-2012} proposed a method to synthesize 3D spin-orbit coupling (SOC) in ultracold atomic systems, analogous to the Rashba effect but extended to three dimensions, termed Weyl SOC. It is demonstrated that this coupling can be realized using Raman transitions connecting four atomic states arranged in a tetrahedral geometry. A key result is the presence of a protected Dirac point in the energy spectrum, which remains robust against uniform Zeeman fields. Our results open avenues for studying the CVLL state with ultracold atoms under experimentally controllable conditions.

Another way to observe the proposed CVLL state is by investigating the magnetic rotons -- excitations with energy minima on a 3D moat-band, discussed in  Ref.~\cite{ST-2004}. The work focuses on the role of magnetic rotons within the Landau-Ginzburg theory of quantum phase transitions. It is shown that in systems with weak spin-orbit coupling and broken inversion symmetry, such as MnSi, paramagnon excitations exhibit a moat-like minimum at a finite wavevector. 
Through self-consistent Hartree and renormalization group (RG) calculations, it is established that the effective action for the collective magnetization includes dynamic and static contributions, with a kinetic term reflecting the moat-like structure. 

Another perspective for the realization of CVLL is the existence of rotons as vortex-antivortex pairs, low-energy collective excitations in Bose-Einstein condensation~\cite{Balents-Fisher} in superfluid $^4$He, which appear with the moat-like dispersion~\cite{Iord-Pit-1980}. The dispersion for rotons mimics the moat dispersion $\epsilon(p)=\Delta+\frac{(p-p_*)^2}{2m_*}$ where $\Delta$ is the roton gap, $p_*$ is the roton minimum momentum, and $m_*$ is the effective roton mass~\cite{Mel}. Such a quasiparticle excitation is the signature of $^4$He in a low-temperature limit. Rotons are also observed in fractional (anomalous) quantum Hall effect both experimentally and numerically~\cite{FAQH-2024, QHE-2022, Oblate}. Rotons in superfluids are also used to detect dark matter by measuring their interaction with the vacuum interface~\cite{Hertel}. Our studies suggest that at low temperatures and at the low-density limit of rotons in $^4$He, the system would exhibit chiral surface states and vortex-line excitations inherent to the CVLL state.

Notably, the 3D moat dispersion also emerges in a spectrum in a dense quark matter, particularly in heavy-ion collisions at low beam energies~\cite{Pisar-2021}. The moat dispersion affects particle production and correlations, leading to distinctive signatures that can be observed experimentally. These include particle distributions and multiparticle correlations that peak at non-zero momenta and differ significantly from typical phases where the lowest energy occurs at zero momentum. The signatures of the predicted CVLL state may become more pronounced and potentially observable because of the preserved U(1) symmetry with the appearance of spatially elongated vortex lines, possibly indicative of the CVLL state of a quantum pion liquid.  This implies that the experimental detection of these spectral features could provide insight into the novel CVLL phases in QCD, potentially observable in facilities such as RHIC's beam energy scan program.

The data that support the findings of this article are openly available\cite{data}.

\section*{Acknowledgment}
The authors are grateful to Devadyouti Das, Alex Kamenev, Nikolay Prokofiev, and Ara Sedrakyan for useful discussions. The authors gratefully acknowledge support from the Simons Center for Geometry and Physics, Stony Brook University, at which some of the research for this paper was performed. The work was partially supported by the Armenian ARPI Remote Laboratory program 24RL-1C024. This work was partially supported by Higher Education and Science Committee of MESCS RA (Research project No. 25PostDoc-1C003).
 
\appendix

\section{The calculation of the polarization operator for cylindrical moat band fermions in (3+1)D} 
\label{polarizationdef}

\vspace{-10pt}
\begin{figure}[ht]
\includegraphics[width=0.23\textwidth]{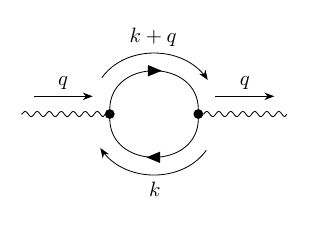}
\includegraphics[width=0.23\textwidth]{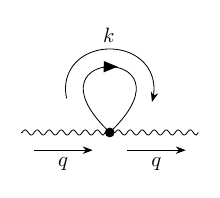}
\caption{One-loop diagrams} 
\label{1loop}
\end{figure}

\begin{figure}[!t]
\includegraphics[width=0.5\textwidth]{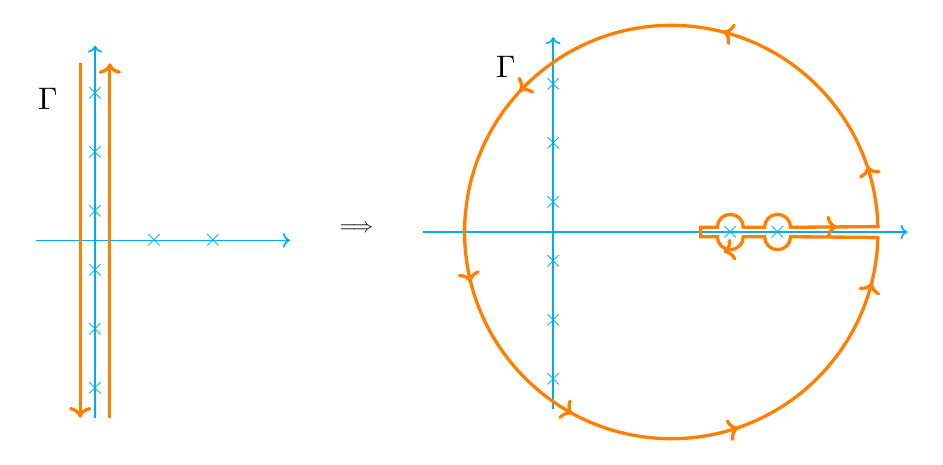}
\caption{Contour $\Gamma$ used in \cref{contour_int}. The left contour can be smoothly deformed into the right one.} 
\label{contour}
\end{figure}

\begin{figure*}[t]
\[
\Pi^{(1,T,\mathrm{RPA})}
=
\vcenter{\hbox{
    \includegraphics[width=0.2\textwidth]{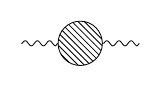}
}}
=
\vcenter{\hbox{
    \includegraphics[width=0.2\textwidth]{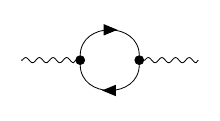}
}}
+
\vcenter{\hbox{
  \includegraphics[width=0.3\textwidth]{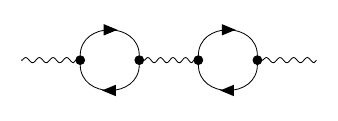}
}}
+ \cdots
\]

\[
\Pi^{(2,T,\mathrm{RPA})}
=
\vcenter{\hbox{%
  \includegraphics[width=0.2\textwidth]{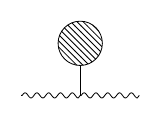}
}}
=
\vcenter{\hbox{%
  \includegraphics[width=0.2\textwidth]{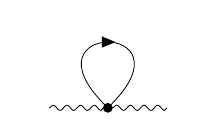}
}}
+
\vcenter{\hbox{%
  \includegraphics[width=0.3\textwidth]{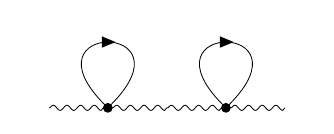}
}}
+ \cdots
\]

\[
\vcenter{\hbox{%
  \includegraphics[width=0.2\textwidth]{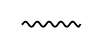}
}}
=
\vcenter{\hbox{%
  \includegraphics[width=0.2\textwidth]{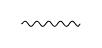}
}}
+
\vcenter{\hbox{
  \includegraphics[width=0.2\textwidth]{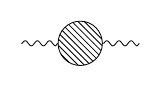}
}}
+
\vcenter{\hbox{%
  \includegraphics[width=0.2\textwidth]{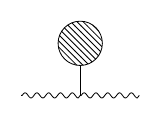}
}}
\]
\caption{RPA diagrams and the Dyson equation. The thick curly line is the gauge field dressed propagator, and the thin curly line is the gauge field bare propagator, $D_{\rm CS}$.} 
\label{RPA-diagram}
\end{figure*}

In this appendix, we demonstrate in detail the calculation of the polarization operator for the cylindrical dispersion to arrive at a low-energy effective theory. 
We start with the effective action Eq.~(\ref{effective}) and keep only the first term
\begin{equation}
    S_{\rm eff} = -\ln \det \left\{ \partial_{\tau}  - \mathcal{A}_0 + K_C(\hat{k}_i - \mathcal{A}_i) \right\}.
\end{equation}
The CS gauge field can be written as the mean field and the fluctuation above the mean field components $\mathcal{A} = \langle \mathcal{A} \rangle + \delta \mathcal{A}$.
At the mean-field level, the $\langle \mathcal{A} \rangle$ effectively creates a magnetic field leading to a fully-filled Landau level and contributes a CS term of level-1. The remaining action can be expanded in the vicinity of
fluctuation beyond the mean-field around $\delta \mathcal{A}$ in the momentum space, giving rise to
\begin{equation}
\begin{aligned}
    S_{\rm eff} =& S_{\rm MF} + S'_{\rm eff} \\
    S_{\rm MF} =& \frac{i}{4\pi} \sum_{\mu,\nu,\rho=\tau,x,y} \epsilon_{\mu\nu\rho} \mathcal{A}_\mu \partial_\nu \mathcal{A}_\rho \\
    S'_{\rm eff} =& -\tr \ln \Gamma
    + \tr \mathcal{A}_{\mu} \frac{\partial \Gamma}{\partial k_{\mu}} \frac{1}{\Gamma} \\
    & + \frac12 \tr \mathcal{A}_{\mu} \frac{\partial\Gamma}{\partial k_{\mu}} \frac{1}{\Gamma} \mathcal{A}_{\nu} \frac{\partial \Gamma}{\partial k_{\nu}} \frac{1}{\Gamma} \\
    & -\frac12 \mathcal{A}_{\mu} \frac{d^2\Gamma}{\partial k_{\mu} \partial k_{\nu}} \frac{1}{\Gamma} \mathcal{A}_{\nu}
    +\cdots,
\end{aligned}
\end{equation}
where we have defined $\Gamma(k_{\mu}) = i\omega + K_C(k_i)$, with $k_{\tau} = i\omega$.
Note that in the action, we abuse the notation a bit using $\mathcal{A}$ for the fluctuation beyond the mean field, which will be adopted in the following appendix.
The first term in $S'_{\rm eff}$ is independent of the gauge field $\mathcal{A}$ and the second term corresponds to a tadpole diagram, which contributes to a constant $\pi \mathcal{A}_0(0)$. Therefore, without loss of generality, we will leave out these terms, giving rise to the effective action
\begin{equation}
    S'_{\rm eff} = \int \frac{d^3\mathbf{q}}{(2 \pi)^3} \mathcal{A}_{\mu}(\mathbf{q}) \Pi_{\mu\nu}(\mathbf{q}) \mathcal{A}_{\nu}(-\mathbf{q}),
\end{equation}
here $\mathbf{q}$ is the transferred momentum and the polarization operator contains two terms $\Pi_{\mu\nu}(\mathbf{q}) = \Pi_{\mu\nu}^{(1)}(\mathbf{q}) + \Pi_{\mu\nu}^{(2)}(\mathbf{q})$ with
\begin{equation}
    \Pi_{\mu\nu}^{(1)}(\mathbf{q}) = \int \frac{d^3\mathbf{k}}{(2\pi)^3} \frac{\partial \Gamma (\mathbf{k})}{\partial k_{\mu}} \frac{1}{\Gamma (\mathbf{k})} \frac{\partial \Gamma (\mathbf{k}+\mathbf{q})}{\partial k_{\nu}} \frac{1}{\Gamma(\mathbf{k}+\mathbf{q})},
\end{equation}
and
\begin{equation}
    \Pi_{\mu\nu}^{(2)}(\mathbf{q}) = -\int \frac{d^3\mathbf{k}}{(2\pi)^3} \frac{\partial^2 \Gamma (\mathbf{k}) }{\partial k_{\mu} \partial k_{\nu}}\frac{1}{\Gamma(\mathbf{k})}.
\end{equation}
where the polarization operator is obtained by performing one-loop self-energy diagram shown in Fig. \ref{1loop} for the infinite cylinder case. $\Pi_{\mu\nu}^{(1)}(\mathbf{q})$ can also be expressed in terms of the free-fermion propagator\cite{AGD}, 

\begin{align}
   \Pi_{\mu\nu}({\bf q})=\int \frac{d^3{\bf k}}{(2\pi)^3}
   j_{\mu} ({\bf k}) \: G ({\bf k})  \: j_{\nu} ({\bf k}+{\bf q})  \: G ({\bf k}+{\bf q})\nn\,, 
\end{align}
where $G$ is the fermion propagator for a given dispersion relation and $j_{\mu}({\bf k})=-\dfrac{\delta G^{-1}({\bf k})}{\delta {\cal A}_{\mu}({\bf k})}$ is the current through the vortex with the momentum $k$ and can be evaluated by taking the derivative of the action $S$ (inverse G) with respect to the fluctuating gauge field ${\cal A}_{\mu}({\bf k})$. In the calculations below, we will only focus on the cylindrical case. One can easily find the current as follows
\begin{align} \label{currentder}
    j_{0}({\bf k}) &= i \\
    j_{\mu}({\bf k}) &= \frac{k_{\mu}}{M}\left(1-\frac{k_0}{k}\right)
\end{align}
where ${\mu}=x,y$ and $k=\sqrt{k_x^2+k_y^2}$.
Consequently, we can anticipate a zero element in the polarization matrix $\Pi_{\mu\nu}({\bf q})$ that is dependent on the $z$-axis. 

The polarization operator is a $\mathbf{k}$ independent constant at zero temperature, therefore, we need to look at the finite temperature case. To compute the finite temperature polarization operator, the integral over the $\omega$ is replaced by a summation over Matsubara frequencies $\omega_n = \frac{(2n+1)\pi}{\beta}$,
\begin{align}
    &\Pi_{\mu\nu}^{(1,T)}(\mathbf{q}) = \\ \nn
    &\int \frac{d^2\mathbf{k}}{(2\pi)^2} \sum_{\omega_n} \frac{1}{\beta} \frac{\partial \Gamma (\mathbf{k})}{\partial k_{\mu}} \frac{1}{\Gamma(\mathbf{k})} \frac{\partial \Gamma (\mathbf{k}+\mathbf{q}) }{\partial k_{\nu}}\frac{1}{\Gamma(\mathbf{k}+\mathbf{q})}
\end{align}
and
\begin{equation}
    \Pi_{\mu\nu}^{(2,T)}(\mathbf{q}) = -\int \frac{d^2\mathbf{k}}{(2\pi)^2} \sum_{\omega_n} \frac{1}{\beta} \frac{\partial^2 \Gamma (\mathbf{K})}{\partial k_{\mu} \partial k_{\nu}} \frac{1}{\Gamma(\mathbf{K})}.
\end{equation}

To begin, let us calculate $\Pi_{\mu\nu}^{(1,T)}$. To perform the summation over the Matsubara frequencies, one can multiply the integrand by $-\frac{\beta}{e^{\beta z} + 1}$, which has poles located at $z = \omega_n$. Then, one can perform the contour integral of $z$ around these poles as shown in \cref{contour}, which leads to the following expression:
\begin{widetext}
\begin{equation} \label{contour_int}
\begin{aligned}
    \Pi_{\mu\nu}^{(1,T)}(\mathbf{q}) &= -\int \frac{d^2\mathbf{k}}{(2\pi)^2} \frac{1}{2\pi i} \oint_{\Gamma} \frac{\partial \Gamma (\mathbf{k})}{\partial k_{\mu}} \frac{1}{\Gamma(\mathbf{k})} \frac{\partial \Gamma(\mathbf{k}+\mathbf{q})}{\partial k_{\nu}} \frac{1}{\Gamma(\mathbf{k}+\mathbf{q})} \frac{1}{e^{\beta z}+1} \\
    &= \int \frac{d^2\mathbf{k}}{(2\pi)^2} \frac{\partial \Gamma(\mathbf{k}) }{\partial k_{\mu}}\frac{\partial \Gamma(\mathbf{k}+\mathbf{q})}{\partial k_{\nu}}
    \left( \left. \frac{1}{\Gamma(\mathbf{k} + \mathbf{q})} \frac{1}{e^{\beta z_1}+1} \right|_{i\omega \to z_1}
    + \left. \frac{1}{\Gamma(\mathbf{k})} \frac{1}{e^{\beta z_2}+1} \right|_{i\omega \to z_2} \right) \\
    &= \int \frac{d^2\mathbf{k}}{(2\pi)^2} \frac{\partial \Gamma(\mathbf{k})}{\partial k_{\mu}} \frac{\partial \Gamma(\mathbf{k}+\mathbf{q})}{\partial k_{\nu}}
    \frac{1}{K_c(\mathbf{k}+\mathbf{q}) - K_c(\mathbf{k})}
    \left( \frac{1}{e^{-\beta K_c(\mathbf{k})} + 1} - \frac{1}{e^{-\beta K_c(\mathbf{k}+\mathbf{q})} + 1} \right).
\end{aligned}
\end{equation}
\end{widetext}
In the second line, the loop around the poles $z=\omega_n$ is deformed to infinity, with the new poles being $z_1 + K_c(k) = 0$ and $z_2 + K_c(k+q) = 0$.

To obtain the low-energy effective theory, we are interested only in the low-energy and momentum limit, i.e., $|\mathbf{q}| \ll k_0$. So, we can proceed with the evaluation by expanding the polarization operator and keeping only the lowest order in $\mathbf{q}$. The resulting polarization operator's components are given by
\begin{equation}
\begin{aligned}
    \Pi_{x\tau}^{(1,T)}(\mathbf{q}) &= \int \frac{d^2\mathbf{k}}{(2\pi)^2} \frac{k_x}{k_x q_x + k_y q_y} \frac{e^{-\beta K_c(\mathbf{k})}}{(e^{-\beta K_c(\mathbf{k})}+1)^2} \\
    &= \int \frac{d^2\mathbf{k}}{(2\pi)^2} \frac{\cos\theta k_x - \sin\theta k_y}{|\mathbf{q}| k_x} \frac{e^{-\beta K_c(\mathbf{k})}}{(e^{-\beta K_c(\mathbf{k})}+1)^2} \\
    &= \frac{q_x}{|\mathbf{q}|^2} \int \frac{d^2\mathbf{k}}{(2\pi)^2} \frac{e^{-\beta K_c(\mathbf{k})}}{(e^{-\beta K_c(\mathbf{k})}+1)^2} \\
    &= \frac{q_x}{|\mathbf{q}|^2} f_{\beta, k_0}.
\end{aligned}
\end{equation}
The second equality is obtained by making a change of variables as:
$k_x \to \cos\theta k_x - \sin\theta k_y$,
$k_y \to \sin\theta k_x + \cos\theta k_y$
with $\tan\theta = \frac{q_y}{q_x}$. In the last line of the equality we defined the integral as follows $f_{\beta, k_0}:= \int \frac{d^2\mathbf{k}}{(2\pi)^2} \frac{e^{-\beta K_c(\mathbf{k})}}{(e^{-\beta K_c(\mathbf{k})}+1)^2} = \int \frac{d^2\mathbf{k}}{(2\pi)^2} \frac{1}{2 \cosh(\beta K_c(\mathbf{k})) + 2}$, which is independent of the transferred momentum $\mathbf{q}$. The integral $f_{\beta, k_0}$ can be further simplified by rescaling $\tilde{\mathbf{k}} = \mathbf{k} / k_0$,
\begin{equation}
    f_{\beta, k_0} = k_0^2 \int \frac{d^2\tilde{\mathbf{k}}}{(2\pi)^2} \frac{1}{2\cosh(\frac{\beta k_0^2}{2M} (|\tilde{\mathbf{k}}| - 1)^2) + 2},
\end{equation}
which is proportional to $k_0^2$ and dependent on a dimensionless parameter $\frac{\beta k_0^2}{2M}$. Fig. \ref{mass} shows the functional behavior of $f_{\beta, k_0}$ versus this parameter. By implementing the same method, one can obtain
\begin{align}
    \Pi_{y\tau}^{(1,T)}(\mathbf{q}) &= \frac{q_y}{|\mathbf{q}|^2} f_{\beta, k_0}, \\
    \Pi_{\tau\tau}^{(1,T)}(\mathbf{q}) &= \beta f_{\beta,k_0}, \\
    \Pi_{ij}^{(1,T)}(\mathbf{q}) &= m_{ij},
\end{align}
where
\begin{equation}
\begin{aligned}
    m_{xx} = m_{yy} &= \frac{\beta k_0^4}{M^2} \int \frac{d^2\tilde{\mathbf{k}}}{(2\pi)^2} \frac{\frac{1}{2} (|\tilde{\mathbf{k}}|-1)^2}{2\cosh(\frac{\beta k_0^2}{2M} (|\tilde{\mathbf{k}}| - 1)^2) + 2}, \\
    m_{xy} = m_{yx} &= \frac{k_0^2}{M} \int \frac{d^2\tilde{\mathbf{k}}}{(2\pi)^2} \frac{\frac{1}{4|\tilde{\mathbf{k}}|} - \frac{3}{4}}{2\cosh(\frac{\beta k_0^2}{2M} (|\tilde{\mathbf{k}}| - 1)^2) + 2}.
\end{aligned}
\end{equation}

Similarly, one can compute the $\Pi_{\mu\nu}^{(2,T)}$ which, up to order $O(q)$, gives rise to
\begin{align}
    \Pi_{\tau\mu}^{(2,T)}(\mathbf{q}) &= \Pi_{xy}^{(2,T)}(\mathbf{q}) = 0, \\
    \Pi_{ii}^{(2,T)}(\mathbf{q}) &= m_i,
\end{align}
where
\begin{equation}
    m_x = m_y = \frac{k_0^2}{M}\int \frac{d^2 \tilde{\mathbf{k}}}{(2\pi)^2} \frac{1}{e^{-\frac{\beta k_0^2}{2M} (|\tilde{\mathbf{k}}| - 1)^2}+1} \left(\frac{1}{2|\tilde{\mathbf{k}}|} - 1\right).
\end{equation}

In summary, the polarization operators can be expressed in matrix form as follows:
\begin{equation}
    \Pi_{\mu\nu}^{(1,T)}(\mathbf{q})
    =
    \begin{pmatrix}
    \beta f_{\beta,k_0} & -f_{\beta,k_0} \frac{q_x}{|\mathbf{q}|^2} & -f_{\beta,k_0} \frac{q_y}{|\mathbf{q}|^2}\\
    f_{\beta,k_0} \frac{q_x}{|\mathbf{q}|^2} & m_{xx} & m_{yx} \\
    f_{\beta,k_0} \frac{q_y}{|\mathbf{q}|^2} & m_{xy} & m_{yy} 
    \end{pmatrix}
\end{equation}
and
\begin{equation}
    \Pi_{\mu\nu}^{(2,T)}(\mathbf{q})
    =
    \begin{pmatrix}
    0 & 0 & 0\\
    0 & m_x & 0 \\
    0 & 0 & m_y 
    \end{pmatrix}.
\end{equation}

At this level, one can derive the equation of motion from the effective action Eq.~(\ref{effective}):
\begin{equation}
    \Pi^{(T)}_{\mu\nu}({\mathbf{q}}) \mathcal{A}_{\nu}({\mathbf{q}}) + \frac{(m+1) L_z}{2} \epsilon^{\mu\rho\sigma}
    F_{\rho\sigma}({\mathbf{q}}) = 0,
\end{equation}
where the field strength tensor is $F_{\mu\nu}({\mathbf{r}}) = \partial_\mu \mathcal{A}_\nu - \partial_\nu \mathcal{A}_\mu$, and 
$\Pi(\mathbf{q})$ is the one-loop polarization operator defined as $\Pi^{(T)}(\mathbf{q}) = \Pi^{(1,T)}(\mathbf{q}) + \Pi^{(2,T)}(\mathbf{q})$. 
This sets up a homogeneous linear equation of $\mathcal{A}$. To have a non-zero solution, the coefficient matrix in front of $\mathcal{A}$, i.e., $\Pi^{(T)}_{\mu\nu}(\mathbf{q}) + (m+1) L_z \epsilon^{\mu\rho\nu} k_{\rho}$, should have zero determinant, which is known as the characteristic equation and, hence, the dispersion relation up to order $O(1/q^2)$ as follows
\begin{equation}
    \omega^2 = \frac{f_{\beta,k_0} (M(\mathbf{q}) + m_x|\mathbf{q}|^2)}
    {L_z^2 (m+1)^2 \beta |\mathbf{q}|^4}.
\end{equation}
where $M({\bf q}) = m_{xx} (q_x^2+q_y^2) - 2 m_{xy} q_x q_y$. To incorporate the CS action and address the singularity in the polarization operators, we perform a random phase approximation (RPA) calculation as shown in Fig. \ref{RPA-diagram}, which leads to
\begin{equation} \label{RPA}
    \Pi_{\mu\nu}^{(a,T,{\rm RPA})}(\mathbf{q})
    = \frac{\Pi_{\mu\nu}^{(a,T)}(\mathbf{q})}{1 - \frac{1}{L_z} D_{\rm CS}(\mathbf{q}) \Pi_{\mu\nu}^{(a,T)}(\mathbf{q})}
\end{equation}
where $a=1,2$ is the index of the two polarization operators and $D_{\rm CS}(\mathbf{q})$ represents CS gauge field propagator.
In imaginary time, the propagator can be explicitly expressed in a matrix form as
\begin{equation}
    D_{\rm CS}(\mathbf{q}) =
    \frac{2\pi}{m+1}
    \begin{pmatrix}
    0 & \frac{q_y}{q^2} & -\frac{q_x}{q^2}\\
    -\frac{q_y}{q^2} & 0 & \frac{q_\tau}{q^2} \\
    \frac{q_x}{q^2} & -\frac{q_\tau}{q^2} & 0
    \end{pmatrix},
\end{equation}
where $m+1$ denotes the level of CS theory and $q^2 = q_{\tau}^2 + q_x^2 + q_y^2$.

The polarization operators can be computed straightforwardly with Eq. (\ref{RPA}):
\begin{widetext}
\begin{equation}
\begin{aligned}
    \Pi_{\mu\nu}^{(1,T,{\rm RPA})}(\mathbf{q}) =
    \frac{f_{\beta,k_0}^2 M({\bf q})}{|{\bf q}|^4 \left( \beta f_{\beta,k_0} M({\bf q}) + (m_{xx}^2-m_{xy}^2) q_{\tau }^2 \right)}
    \begin{pmatrix}
        q_{\tau }^2 & q_{\tau } q_x & q_{\tau } q_y \\
        q_{\tau } q_x & q_x^2 & q_x q_y \\
        q_{\tau } q_y & q_x q_y & q_y^2
    \end{pmatrix},
\end{aligned}
\end{equation}
\end{widetext}
and
\begin{equation}
\begin{aligned}
    \Pi_{\mu\nu}^{(2,T,{\rm RPA})}(\mathbf{q}) =
    \frac{(m+1) L_z}{2\pi}
    \begin{pmatrix}
        0 & 0 & 0 \\
        0 & 0 & \frac{q^2}{q_{\tau}} \\
        0 & -\frac{q^2}{q_{\tau}} & 0
    \end{pmatrix}.
\end{aligned}
\end{equation}
In the last equality, we utilize the divergence of $m_x$, which renders $\Pi_{\mu\nu}^{(2,T,{\rm RPA})}(\mathbf{q})$ purely off-diagonal. These polarization operators reproduce the results presented in the main text Eq. (\ref{Pi}).

The dispersion of the gauge field can be determined by the characteristic equation. At zero temperature, $\Pi^{1,T, {\rm RPA}}=0$, leading to a dispersion for the $ x $- and $ y $-components of the gauge field, forming a light cone: 
\begin{equation} \label{lightcone}
\omega = \pm\sqrt{q_x^2 + q_y^2}.
\end{equation}
At finite temperature, with the help of the matrix determinant lemma,
\begin{equation}
    \det(A + uv^{\rm T}) = \det(A) + v^{\rm T} \mathrm{adj}(A) u,
\end{equation}
and considering that $\Pi^{1,T,{\rm RPA}}$ can be written in a tensor product of a vector with itself, one can compute the full determinant of $\Pi^{T,{\rm RPA}}$ which gives rise to the same light cone dispersion as written in Eq.~(\ref{lightcone}).

\section{The derivation of the boundary action in the low-energy effective field theory}

\label{boundary}

In this appendix, we will consider the second,  ``boundary" term of the effective low-energy action Eq.~(\ref{effective1}) of the main text and show that it is topological, corresponding to a novel action in a reduced spatial dimension defined on a boundary of the system. Using $\partial _{i^\prime}^2=\partial_{i^\prime}\delta_{i^\prime j^\prime}\partial_{j^\prime}=-\varepsilon_{i^\prime l}\varepsilon_{l j^\prime}\partial_{i^\prime}\partial_{j^\prime}$,where the indices correspond to the spatial coordinates $x$ and $y$ and $\varepsilon_{ij}$ is the antisymmetric tensor, one finds
\begin{eqnarray}
\label{bound1}
   && S_{\text{bound}}=\frac{1}{2}\int dt\; \text{sgn}(t)\int dx dy \;\epsilon_{ij} \delta\mathcal{A}_i  \nabla^2
    \delta\mathcal{A}_j,\nonumber\\
  &&  =-\frac{1}{2} \int dt\;\text{sgn}(t) \int dx dy \;\left\{\varepsilon_{i^\prime l}\partial_{i^\prime}\left[\varepsilon_{ij}\delta\mathcal{A}_i\varepsilon_{lj^\prime}\partial_{j^\prime}\delta\mathcal{A}_j\right]\right.\nonumber\\
   && \left.-\varepsilon_{i^\prime l}(\partial_{i^\prime}\delta\mathcal{A}_i)(\partial_{j^\prime}\delta\mathcal{A}_j)\varepsilon_{ij}\varepsilon_{lj^\prime}\right\},
\end{eqnarray}
where the sum over repeating indices is assumed. The second term in this expression, $\varepsilon_{i^\prime l}(\partial_{i^\prime}\delta\mathcal{A}_i)(\partial_{j^\prime}\delta\mathcal{A}_j)\varepsilon_{ij}\varepsilon_{lj^\prime}=\varepsilon_{ij}(\partial_{i^\prime}\delta\mathcal{A}_i)(\partial_{j^\prime}\delta\mathcal{A}_j)\delta_{i^\prime j^\prime}$, is vanishing due to the antisymmetric property of $\varepsilon_{ij}$. Therefore, upon introducing a {\em dual} vector field $\tilde{\mathcal{A}}_l\equiv\varepsilon_{ij}\delta\mathcal{A}_i\varepsilon_{lj^\prime}\partial_{j^\prime}\delta\mathcal{A}_j$, one can rewrite the 
action Eq.~(\ref{bound1}) as
\begin{eqnarray}
\label{bound2}
S_{\text{bound}}&&=-\frac{1}{2}\int dt\; \text{sgn}(t)\int dx dy \;\epsilon_{i^\prime l} \partial_{i^\prime} \tilde{\mathcal{A}}_l\nonumber\\
&&=-\frac{1}{2}\int dt\; \text{sgn}(t)\oint dr\;\varepsilon^{lk}n_k\tilde{\mathcal{A}}_l, 
\end{eqnarray}
where we used Stokes' theorem relating the integral of the curl of the auxiliary vector field $\tilde{\mathcal{A}}_l$ over the 2D surface to the line integral around the boundary of the system. Here, ${\bf n}=(n_x,n_y)$ is the unit vector normal to the 1D boundary of the 2D system at a fixed $z$ in 3D real space. Using the fact that $dr\;\varepsilon^{lk}n_k\varepsilon_{lj^\prime}=dr\;n_{j^\prime},$ one can rewrite Eq.~(\ref{bound2}) in the following form
\begin{equation}
    S_{\text{bound}}=-\frac{1}{2}\int dt\; \text{sgn}(t)\oint dr\;n^l \varepsilon_{ij}\delta{\mathcal{A}}_i\partial_{l}\delta\mathcal{A}_j.
\end{equation}
This reproduces the boundary term in the low-energy effective action Eq.~(\ref{effbound}) of the main text. 

Alternatively, using the identity $\varepsilon_{ij}\varepsilon_{lj^\prime}=\delta_{il}\delta_{jj^\prime}-\delta_{ij^\prime}\delta_{jl}$, one finds that 
\begin{eqnarray}
\tilde{\mathcal{A}}_l=\delta{\mathcal{A}}_l(\partial_{j}\delta\mathcal{A}_j)-\delta{\mathcal{A}}_j(\partial_{j}\delta\mathcal{A}_l). 
\end{eqnarray}
Substituting this expression into Eq.~(\ref{bound2}) and integrating by parts, one finds an alternative form of the boundary action:
\begin{equation}
    S_{\text{bound}}=-\int dt\; \text{sgn}(t)\oint dr\;\varepsilon^{lk}n_k\delta\mathcal{A}_l(\partial_{j}\delta\mathcal{A}_j).
\end{equation}
Note that the coefficient $1/2$ in Eq.~(\ref{bound2}) is now canceled in this expression.

\end{document}